\documentclass[reprint,nofootinbib,showpacs]{revtex4}
\usepackage{graphicx}  
\usepackage{dcolumn}   
\usepackage{bm}        
\usepackage{amssymb}
\usepackage{hyperref}
\usepackage{multirow}

\begin{document}

\title{Probing new physics in $B_s \to (K,K^*)\tau \nu$ and $B \to \pi \tau \nu$ decays}

\author{N Rajeev${}^{}$}
\email{rajeev@rs.phy.student.nits.ac.in}
\author{Rupak~Dutta${}^{}$}
\email{rupak@phy.nits.ac.in}
\affiliation{
${}$National Institute of Technology Silchar, Silchar 788010, India\\
}

\begin{abstract}
Motivated by the anomalies present in $b \to u$ and $b \to c$ semileptonic decays, we study the
corresponding $B_s \to (K,K^*) \tau \nu$ and $B \to \pi \tau \nu$ decays within an effective field theory formalism.
Our analysis is based on a strict model dependent assumption, i.e., we assume that $b \to u$ and $b \to c$
transition decays exhibit similar new physics pattern. We give prediction of various observables
such as the branching fraction, ratio of branching ratio, lepton side forward-backward asymmetry,
longitudinal polarization fraction of the charged lepton and convexity parameter in the standard
model and in the presence of vector type new physics couplings.
\end{abstract}

\maketitle

\section{Introduction}
Study of lepton flavor non-universality in the $B$ meson systems have been the center of interest both
theoretically and experimentally over the last decade.  
Disagreement between the SM expectations and the experimental measurements (BaBar, Belle and LHCb) in $B \to D^{(*)} l \nu$
and $B_c \to J/\Psi l \nu$ undergoing $b \to (c,u) l \nu$ quark level transitions are well reflected in the flavor ratios $R_D$, $R_{D^{\ast}}$
and $R_{J/\Psi}$ defined as,

\begin{eqnarray}
R_{D^{(*)}} = \frac{\mathcal{B}(B \to D^{(*)} \tau \nu)}{\mathcal{B}(B \to D^{(*)} l \nu)}\,, \qquad
R_{J/\Psi} =\frac{\mathcal{B}(B_c \to J/\Psi \tau \nu)}{\mathcal{B}(B_c \to J/\Psi l \nu)} \nonumber
\end{eqnarray}

\begin{table}[ht!]
\centering
{\begin{tabular}{|c|c|c|}
    \hline
Ratio of branching ratio&SM prediction&Experimental prediction  \\
    \hline
    \hline
    $R_D$ & $0.300 \pm 0.008$~\cite{Lattice:2015rga,Na:2015kha,Aoki:2016frl,Bigi:2016mdz} & $0.407 \pm 0.039 \pm 0.024$~\cite{Lees:2013uzd,Huschle:2015rga,Sato:2016svk,Hirose:2016wfn,Aaij:2015yra}  \\
    \hline
    $R_{D^{\ast}}$ & $0.258 \pm 0.005$~\cite{Fajfer:2012vx,Bernlochner:2017jka,Bigi:2017jbd,Jaiswal:2017rve} & $0.304 \pm 0.013 \pm 0.007$~\cite{Lees:2013uzd,Huschle:2015rga,Sato:2016svk,Hirose:2016wfn,Aaij:2015yra} \\
    \hline
    $R_{J/\Psi}$ & $[0.20,\,0.39]$~\cite{Cohen:2018dgz} & $0.71 \pm 0.17 \pm 0.18$~\cite{Aaij:2017tyk} \\
    \hline
    $\mathcal{B}(B \to \tau \nu)$ & $(0.84 \pm 0.11)\times 10^{-4}$~\cite{Bona:2009cj} & $(1.09 \pm 2.4)\times 10^{-4}$~\cite{Bernlochner:2015mya} \\
    \hline
    $R_{\pi}^l$ &  $0.566$ & $0.698 \pm 0.155$~\cite{Patrignani:2016xqp} \\
    \hline
    $R_{\pi}$ &  $0.641$~\cite{Patrignani:2016xqp} &  $< 1.784$~\cite{Bernlochner:2015mya} \\
    \hline
\end{tabular}}
\caption{The SM prediction and the world averages of the ratio of branching ratios for various decay modes}
\label{avg}
\end{table}

In Table~\ref{avg}, we report the precise SM predictions and the experimental measurements of the various decay modes. 
The combined deviation of $3.78\sigma$ in $R_D$ and $R_{D^*}$ and around $1.3\sigma$ in $R_{J/\Psi}$ from SM expectation is observed.
Similarly, the average value of the branching ratio $\mathcal{B}(B \to \tau \nu)$ reported by BaBar and Belle experiments is not in good 
agreement with the SM expectations.
Although, the $\mathcal{B}(B \to \pi l \nu)$ is consistent with the SM, the ratio 
$R_{\pi}^l = \left(\tau_{B^0}/\tau_{B^-}\right) {\mathcal{B}(B \to \tau \nu)}/$ ${\mathcal{B}(B \to \pi l \nu)}$ shows mild deviation. 
Similar deviations are also observed in the ratio $R_{\pi} = {\mathcal B(B \to \pi\tau\nu)}/{\mathcal B(B \to \pi\,l\,\nu)}$ as well. 
Motivated by these anomalies, we study the implications of $R_D$, $R_{D^{\ast}}$, $R_{J/\Psi}$, and $R_{\pi}^l$ anomalies 
on $B_s \to (K,\,K^{\ast})\tau\nu$ and $B \to \pi\tau\nu$ semileptonic decays in a model dependent way.

\section{Theory}

\subsection{Effective Lagrangian}
The effective Lagrangian for $b \to u\,l\,\nu$ transition decays in the presence of vector type NP couplings is 
of the form~\cite{Dutta:2013qaa}
\begin{eqnarray}
\label{effl}
\mathcal L_{\rm eff} &=&
-\frac{4\,G_F}{\sqrt{2}}\,V_{u b}\,\Bigg\{(1 + V_L)\,\bar{l}_L\,\gamma_{\mu}\,\nu_L\,\bar{c}_L\,\gamma^{\mu}\,b_L +
V_R\,\bar{l}_L\,\gamma_{\mu}\,\nu_L\,\bar{c}_R\,\gamma^{\mu}\,b_R \nonumber \\
&&+
\widetilde{V}_L\,\bar{l}_R\,\gamma_{\mu}\,\nu_R\,\bar{c}_L\,\gamma^{\mu}\,b_L 
+
\widetilde{V}_R\,\bar{l}_R\,\gamma_{\mu}\,\nu_R\,\bar{c}_R\,\gamma^{\mu}\,b_R \Bigg\} + {\rm h.c.}\,,
\end{eqnarray}
where, $G_F$ is the Fermi coupling constant and $|V_{ub}|$ is the CKM matrix element. 
$V_L$, $V_R$ are the NP Wilson coefficients~(WCs) involving left-handed neutrinos, and the WCs referring to tilde terms 
involve right-handed neutrinos.

Using the effective Lagrangian, we calculate the three body differential decay distribution for the $B \to (P,\,V)\,l\,\nu$ decays. 
The final expressions pertaining to the psudoscalar and vector differential decay rates can be found in~\cite{Rajeev:2018txm}. 

In general, we define the ratio of branching ratio as
\begin{eqnarray}
R = \frac{\mathcal B(B_q \to M\,\tau\,\nu)}{\mathcal B(B_q \to M\,l\,\nu)}\,,
\end{eqnarray}
where $M = K,\,K^{\ast},\,\pi$ and $l=\mu$.
We also define various $q^2$ dependent observables such as differential branching ratio $DBR(q^2)$, ratio of branching ratio $R(q^2)$, 
forward backward asymmetry $A_{FB}^l(q^2)$, polarization fraction of the charged lepton $P^{l}(q^2)$ and convexity parameter $C_{F}^l (q^2)$ 
for the decay modes. For details one can refer to~\cite{Rajeev:2018txm}.

\section{Results and discussion}

\subsection{Standard model predictions}

\begin{figure}[ht!]
\centering
\includegraphics[width=3.6cm,height=2.3cm]{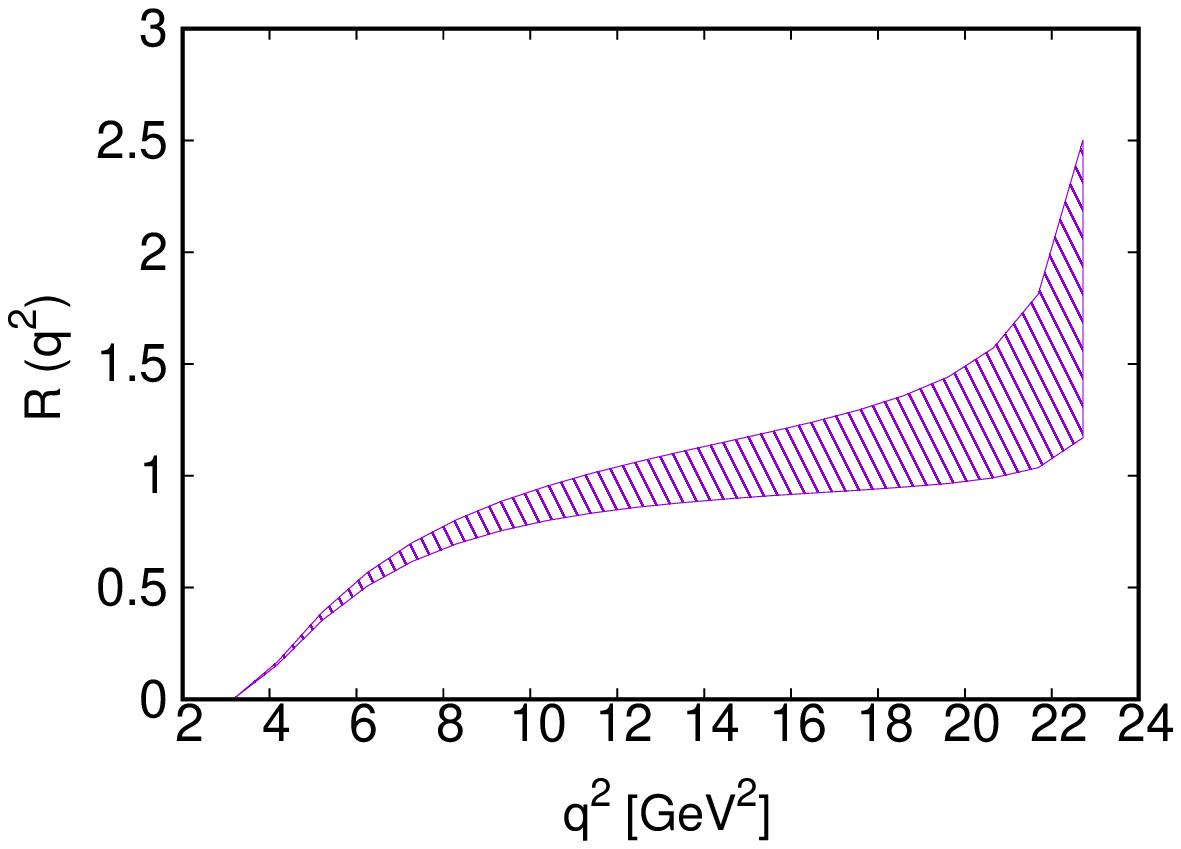}
\includegraphics[width=3.6cm,height=2.3cm]{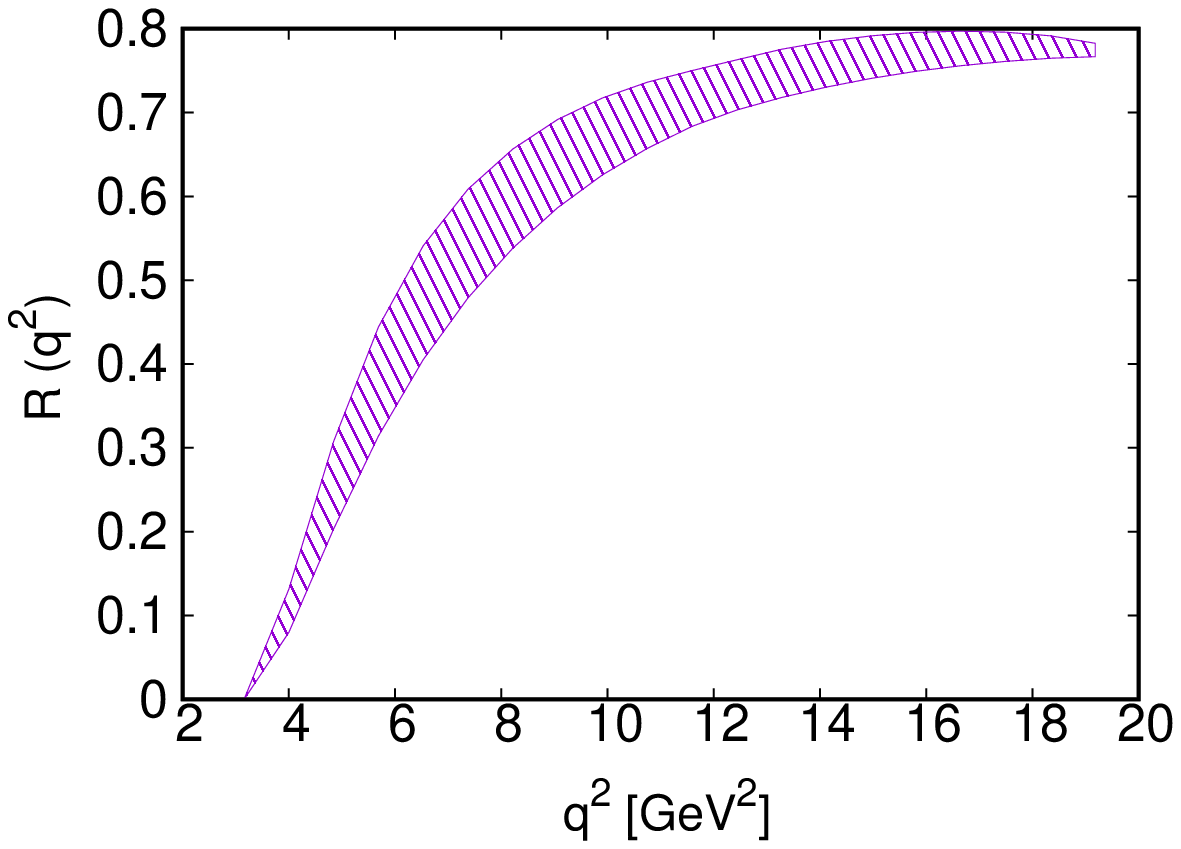}
\includegraphics[width=3.6cm,height=2.3cm]{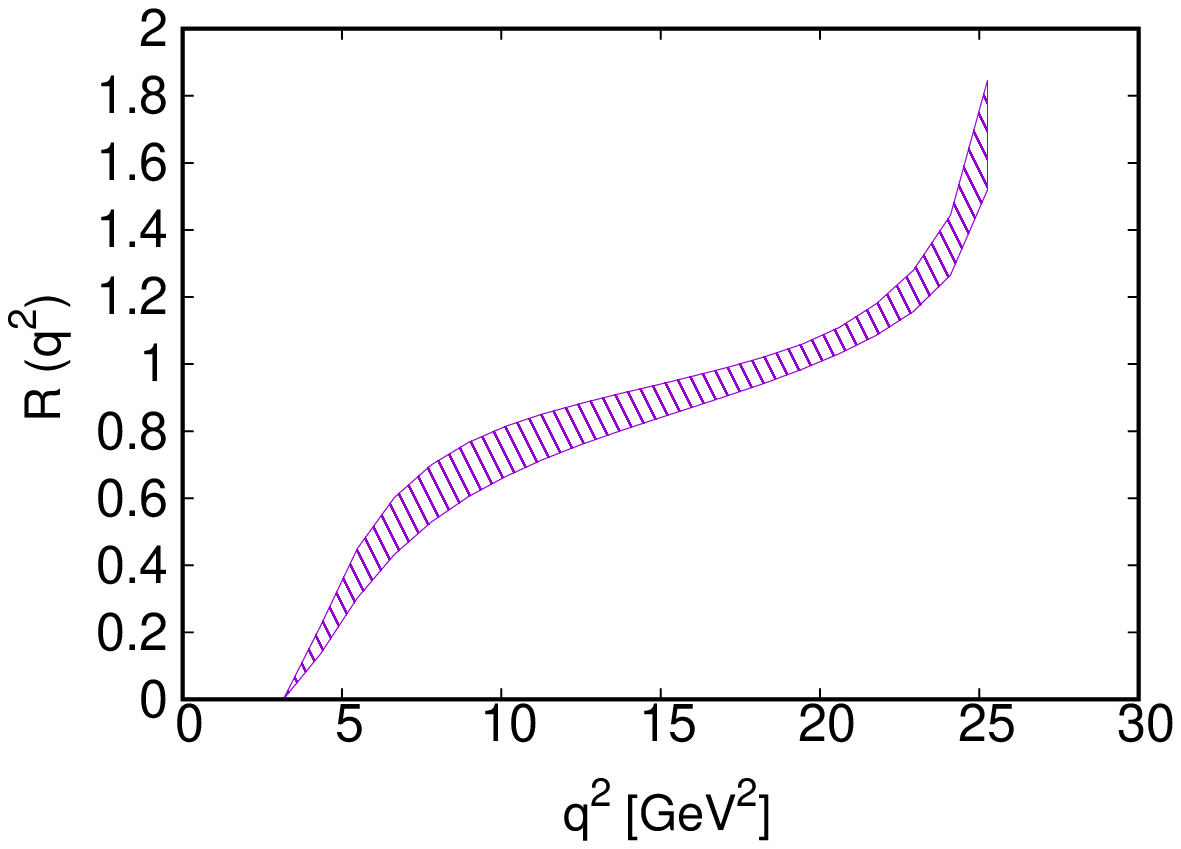}
\includegraphics[width=3.6cm,height=2.3cm]{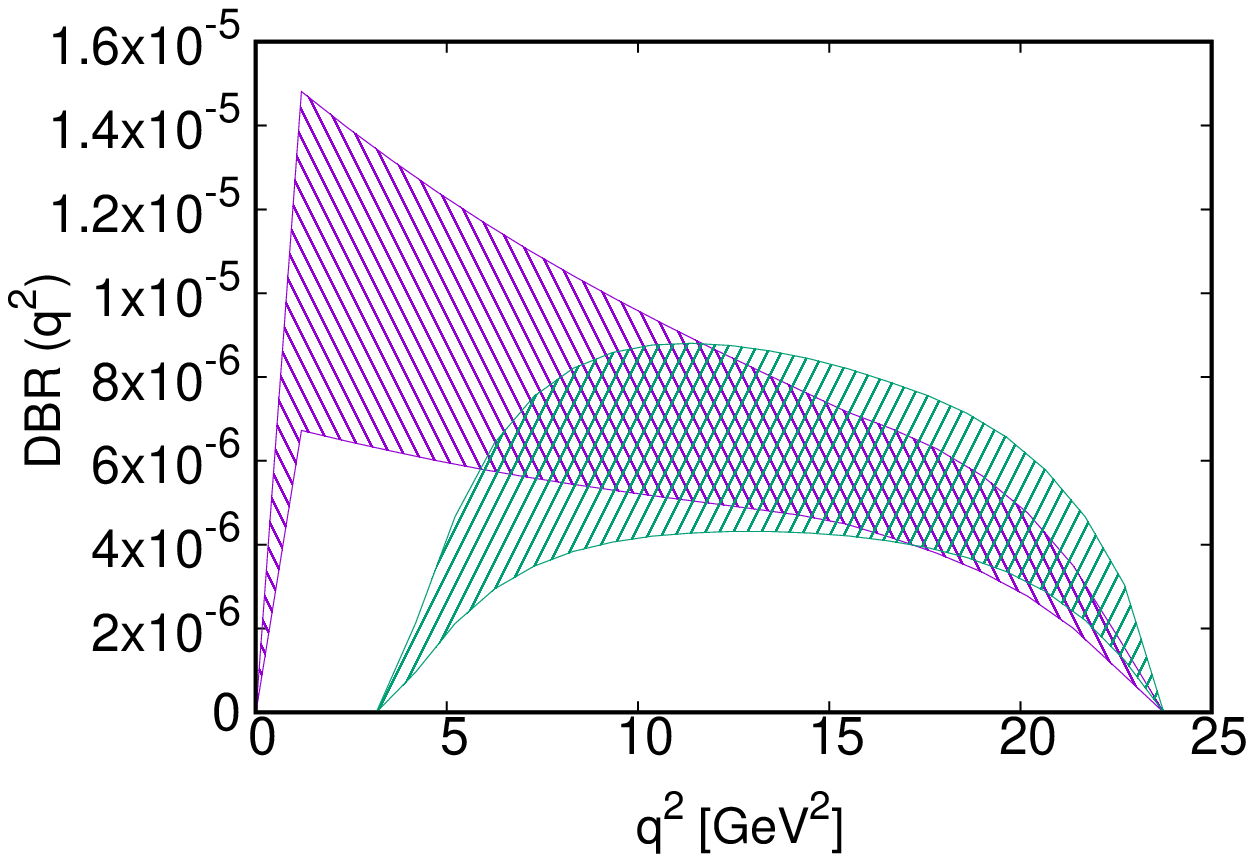}
\includegraphics[width=3.6cm,height=2.3cm]{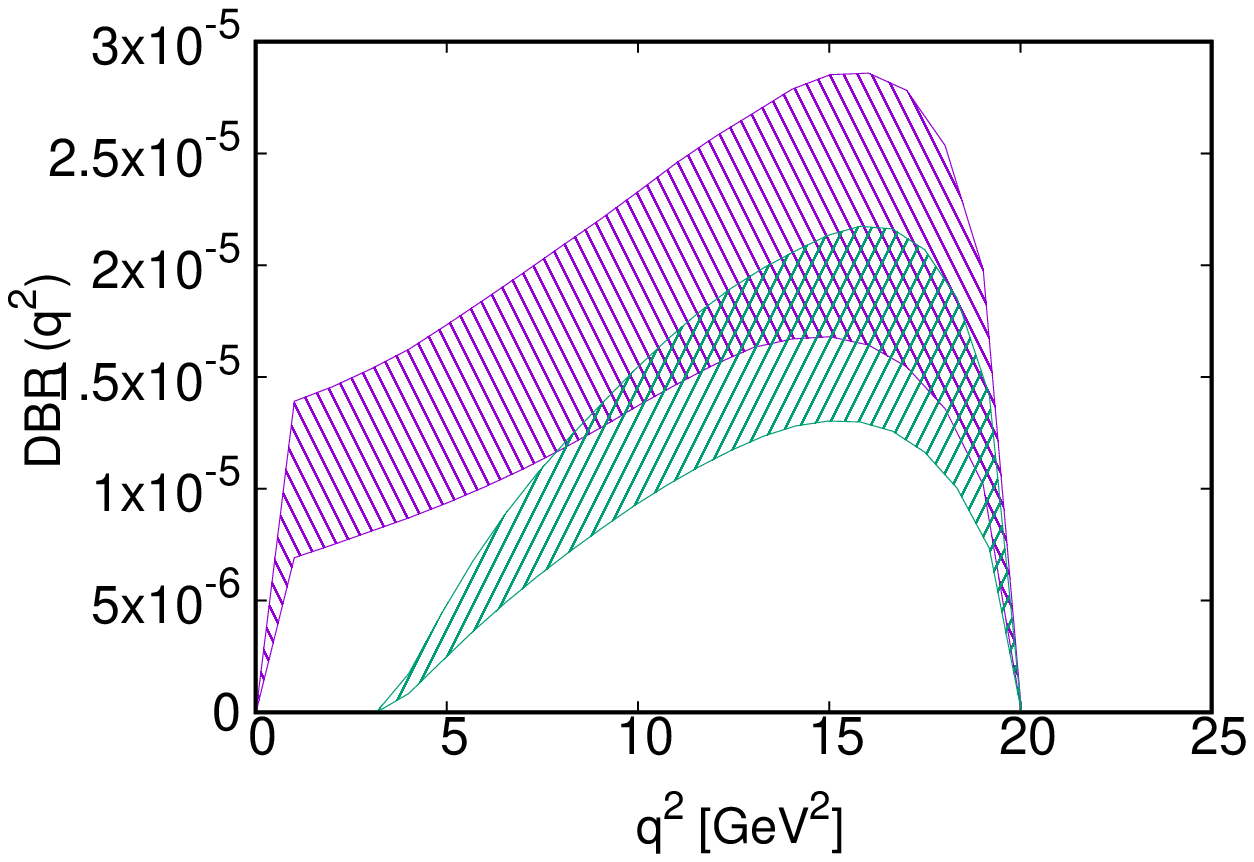}
\includegraphics[width=3.6cm,height=2.3cm]{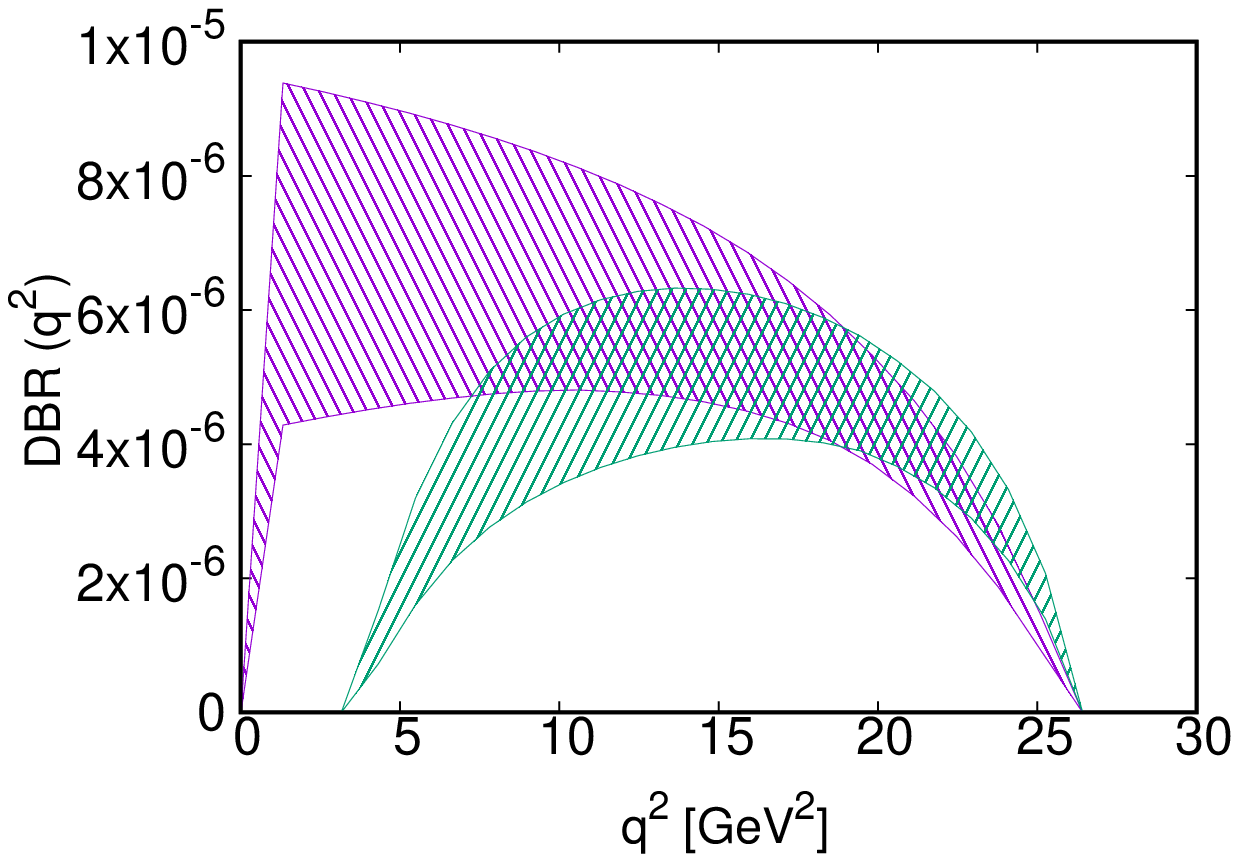}
\includegraphics[width=3.6cm,height=2.3cm]{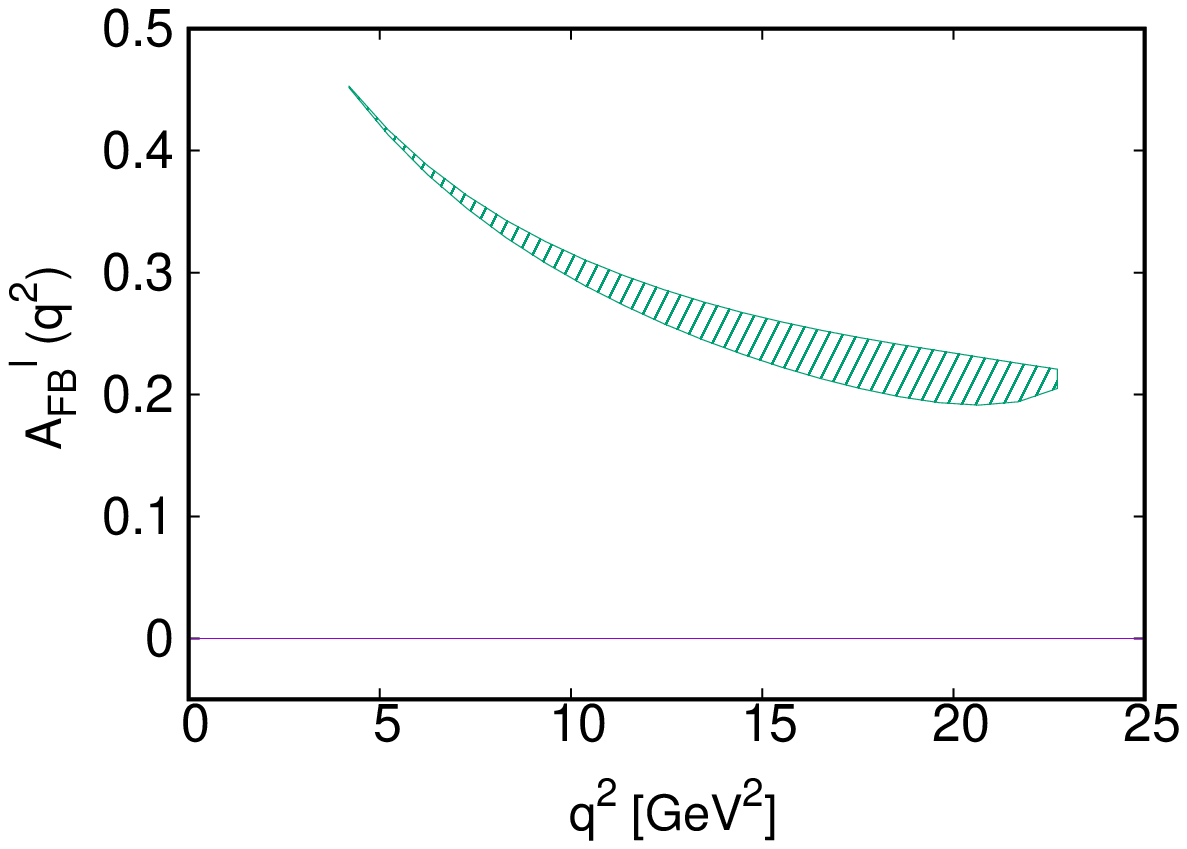}
\includegraphics[width=3.6cm,height=2.3cm]{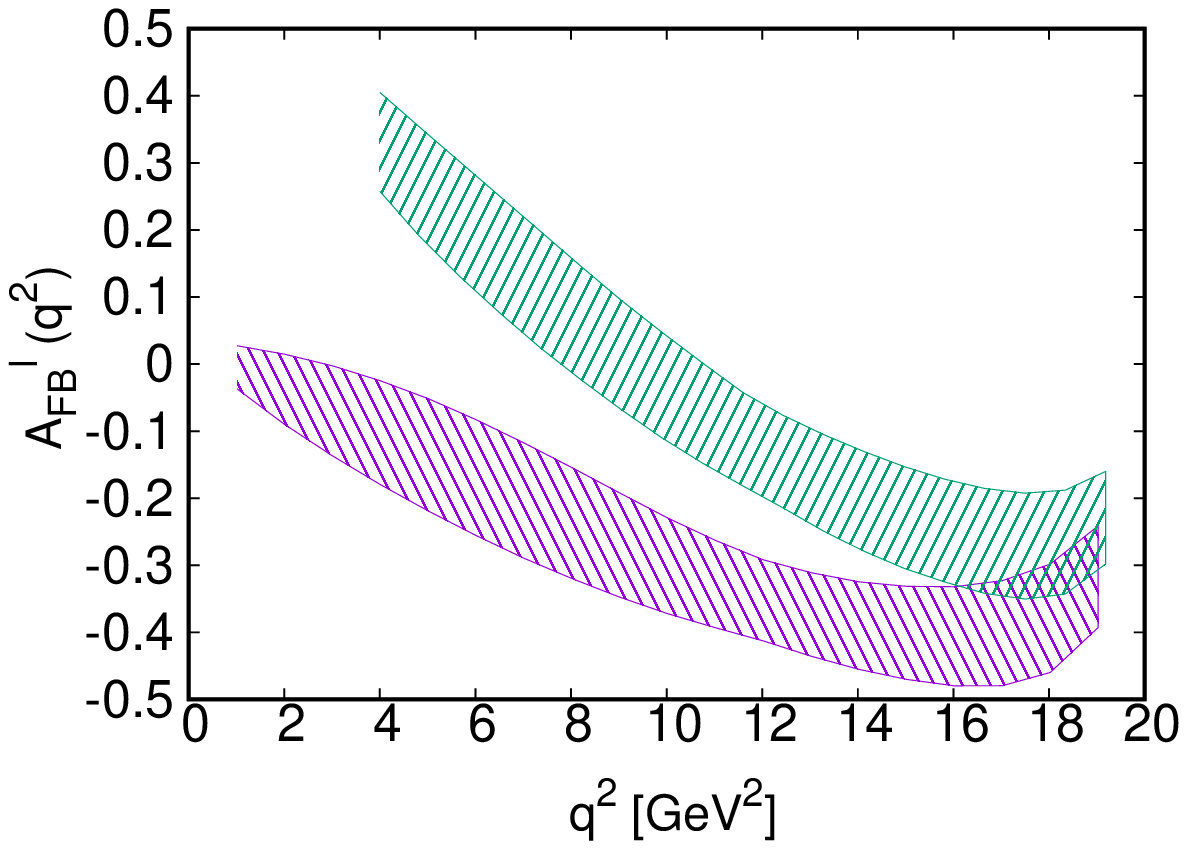}
\includegraphics[width=3.6cm,height=2.3cm]{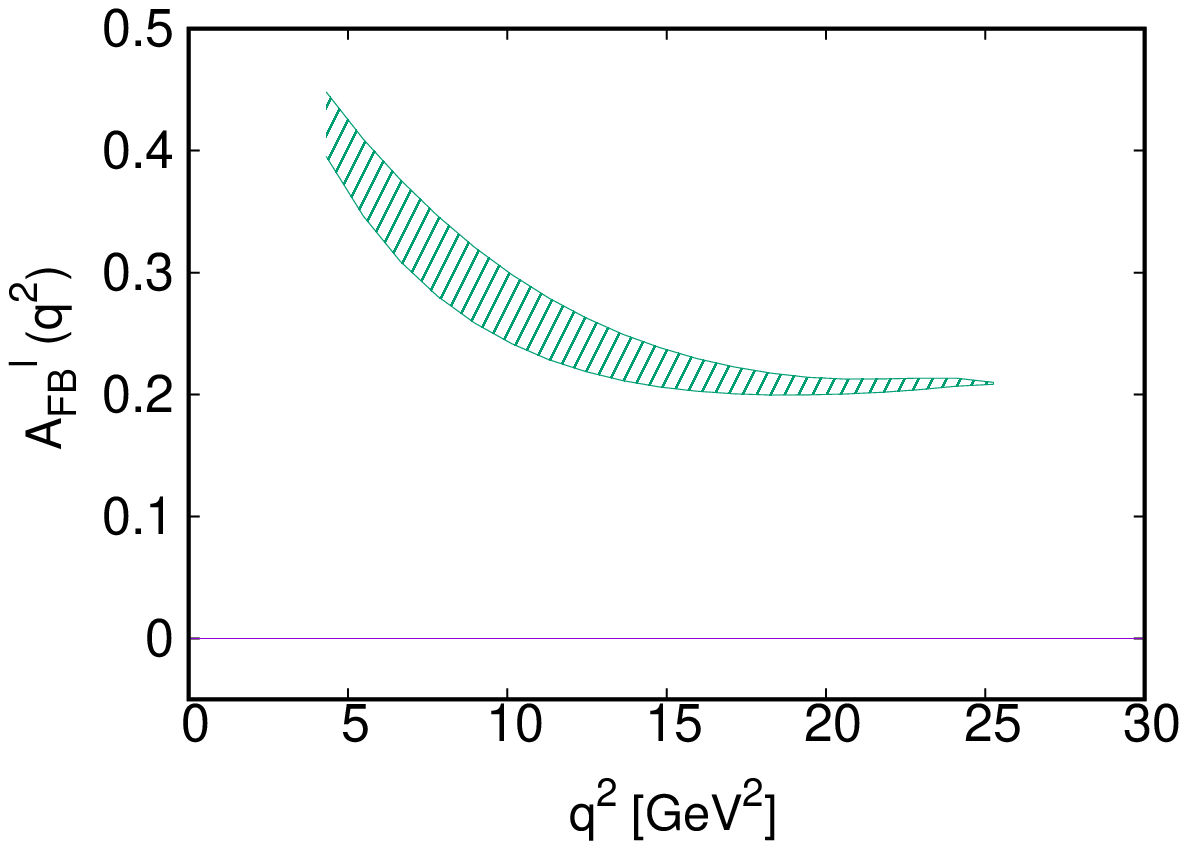}
\includegraphics[width=3.6cm,height=2.3cm]{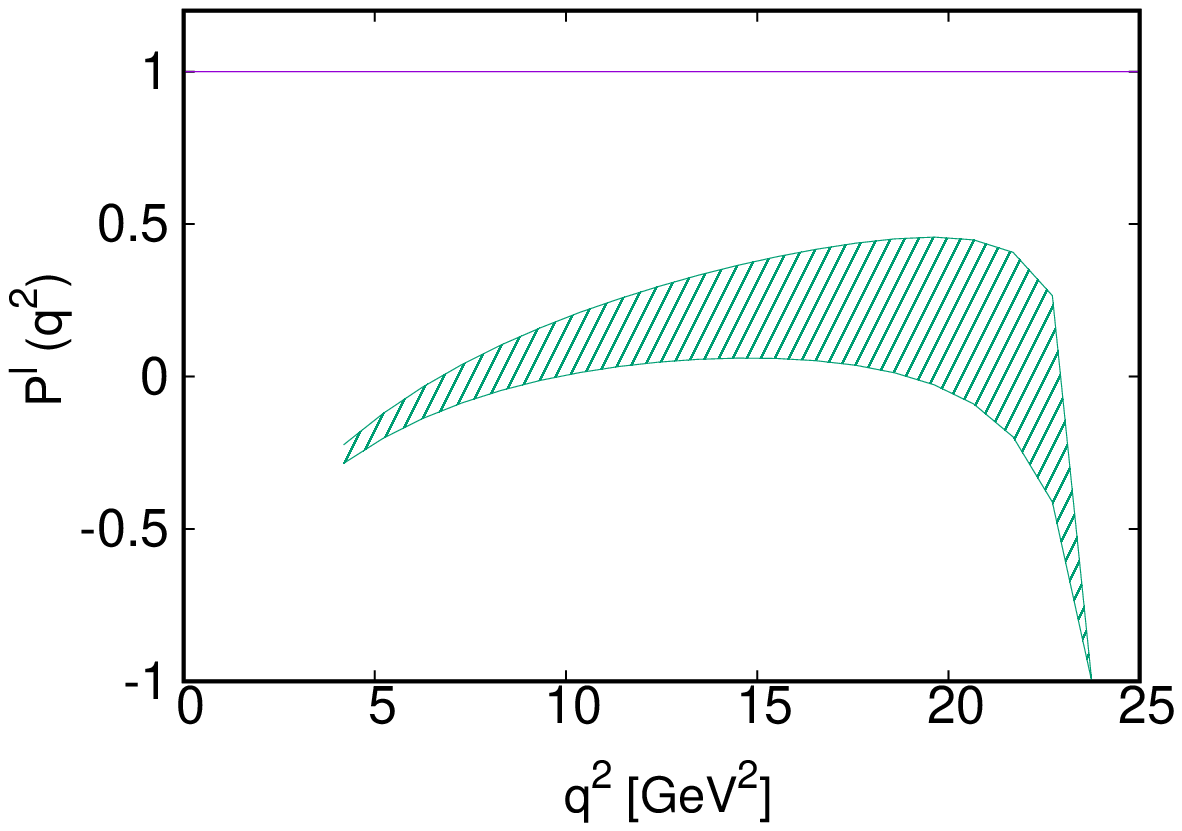}
\includegraphics[width=3.6cm,height=2.3cm]{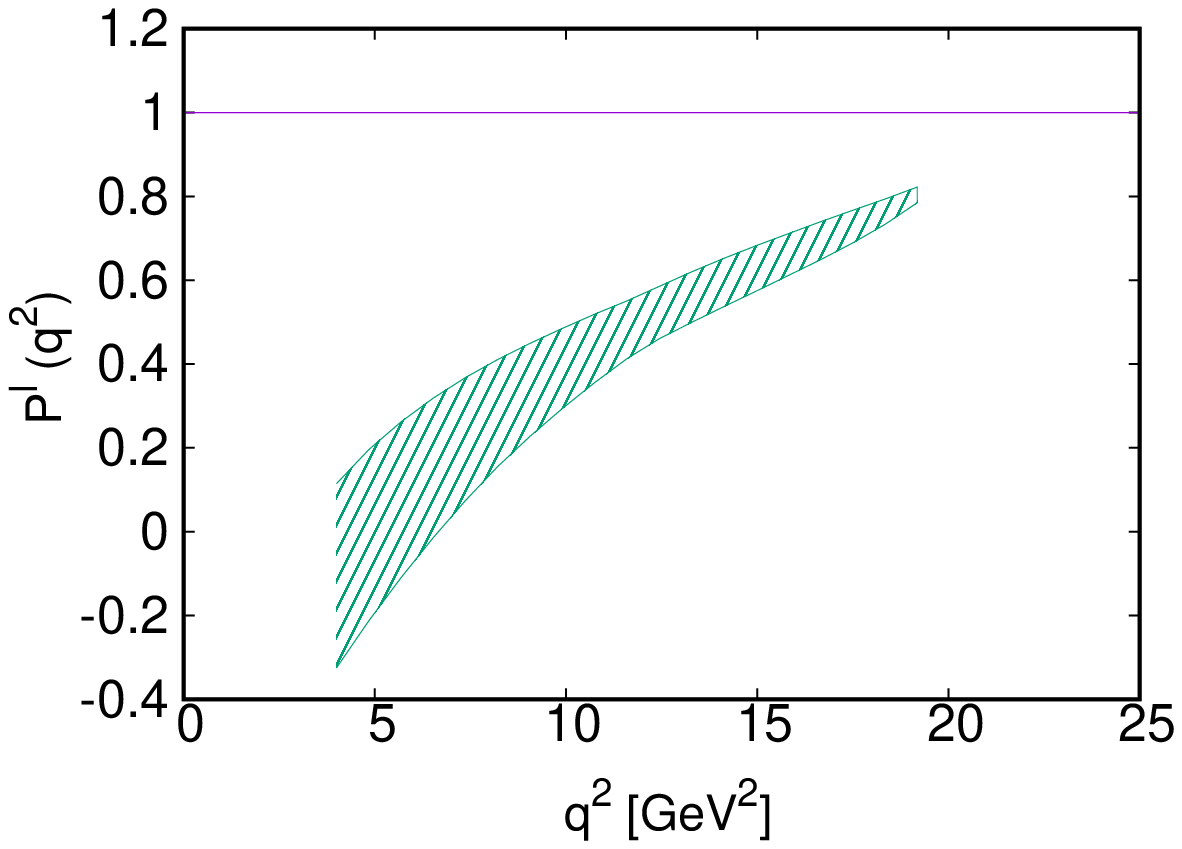}
\includegraphics[width=3.6cm,height=2.3cm]{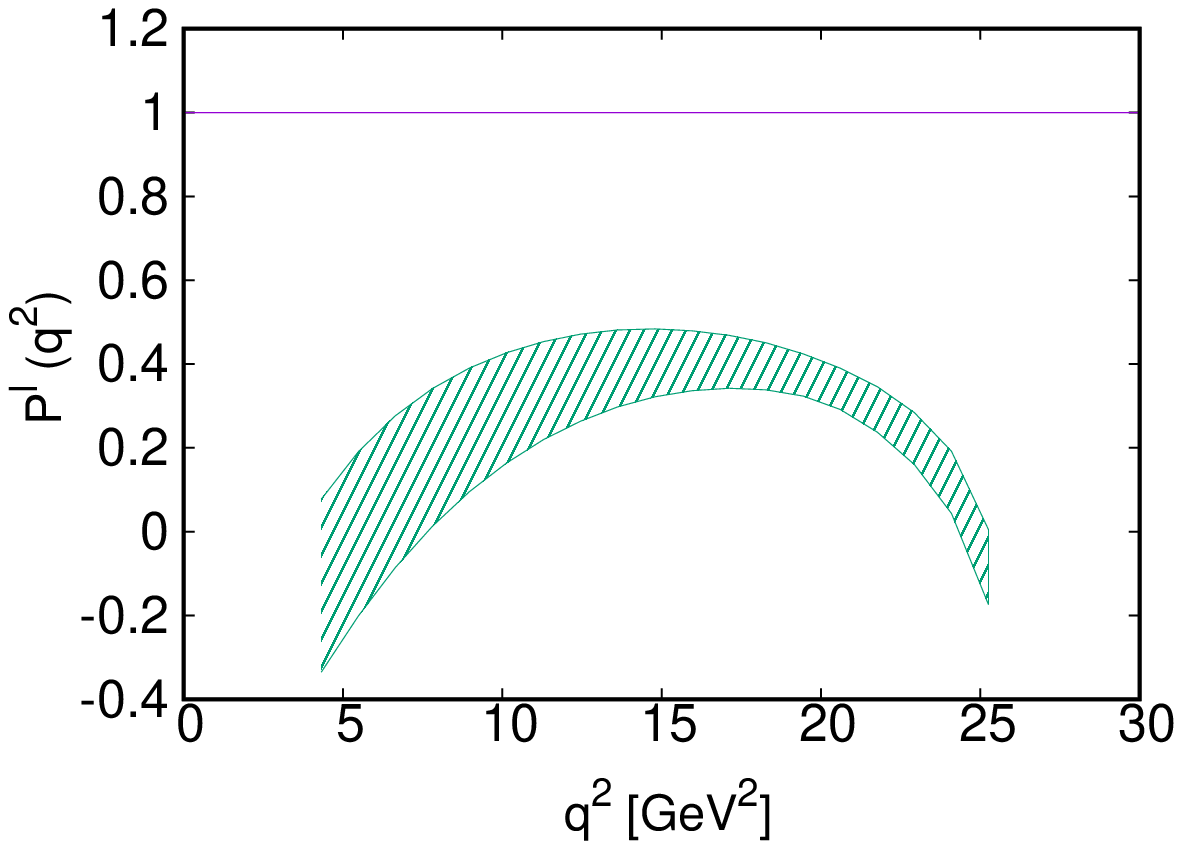}
\includegraphics[width=3.6cm,height=2.3cm]{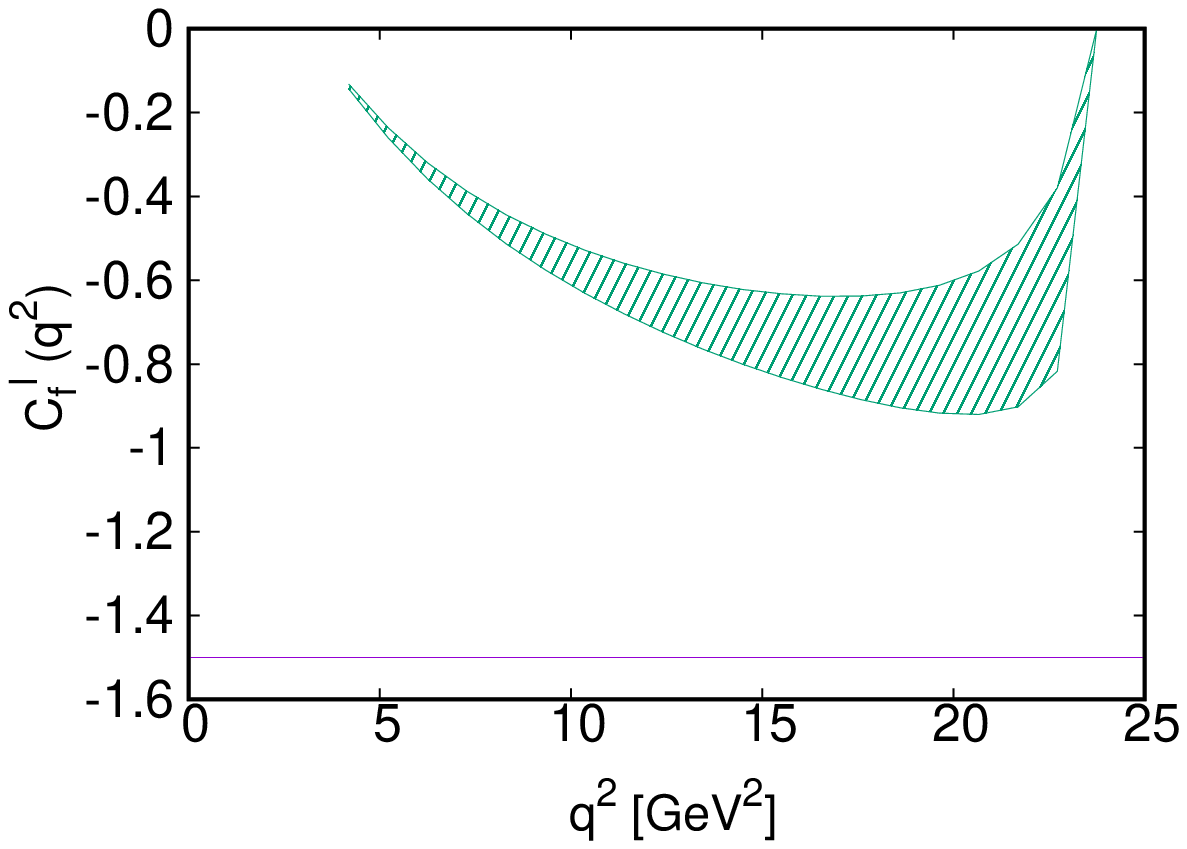}
\includegraphics[width=3.6cm,height=2.3cm]{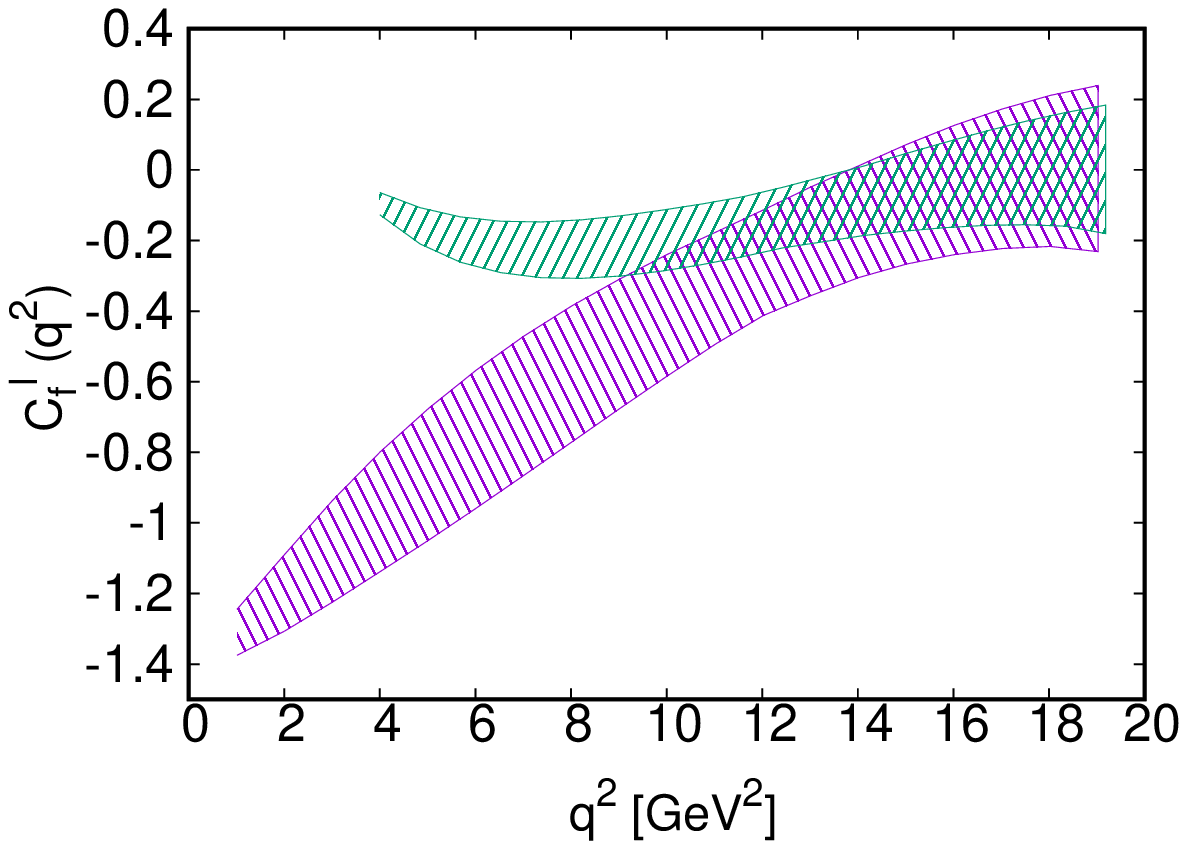}
\includegraphics[width=3.6cm,height=2.3cm]{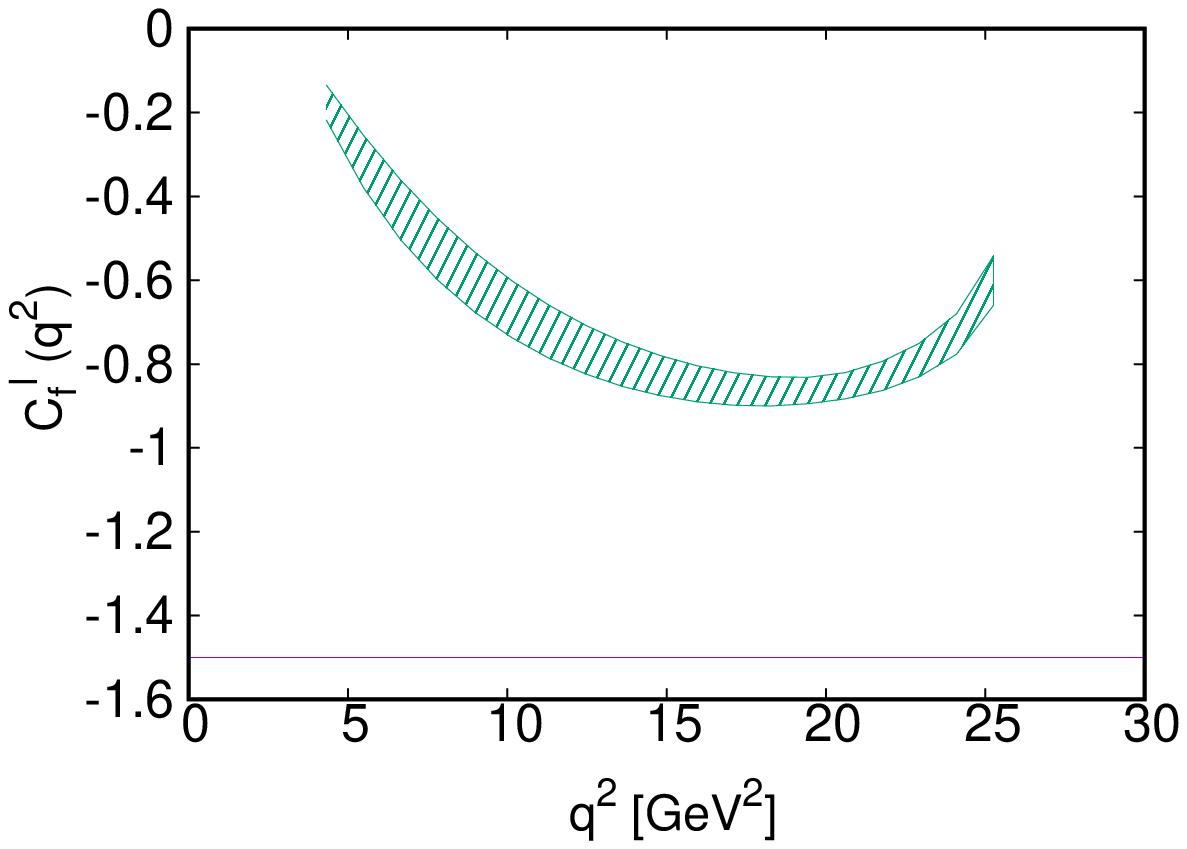}
\caption{$q^2$ dependent observables of $B_s \to K\,l\,\nu$~(first column), $B_s \to K^{\ast}\,l\,\nu$~(second column) and 
$B \to \pi\,l\,\nu$~(third column) decays in the SM for the $\mu$~(violet) and $\tau$~(green) modes.}
\label{figsm}
\end{figure}

The SM central values are reported in Table~\ref{1sigval}. We calculate the central values by considering the central values of the input parameters. 
For the $1\sigma$ ranges, we perform a random scan over the theoretical inputs such as CKM matrix elements and the form factor inputs within
$1\sigma$ of their central values. The significant difference in the $\mu$ mode and the $\tau$ mode are observed. 
The branching ratio of the order of $10^{-4}$ is observed in all the decay modes. The results pertaining $\langle P^{l} \rangle$ and 
$\langle C_{F}^{l} \rangle$ are calculated for the first time for these decay modes.
In Fig.~\ref{figsm}, we show the $q^2$ dependency of all the observables for the $\mu$ mode and the $\tau$ mode.

\begin{table}[ht!]
\centering
\resizebox{\textwidth}{!}{\begin{tabular}{|c|c||c|c|c|c|c||c|}
    \hline
    $B_s \to K l \nu$&  
    &$BR \times 10^{-4}$ &$\langle A_{FB}^{l} \rangle$& $\langle P^{l} \rangle$&$\langle C_{F}^{l} \rangle$&$ R_{{B_s}K}$ \\
    \hline
    \hline
    \multirow{2}{*}{$\mu$ mode}
    & Central value & 1.520 & $6.647\times 10^{-3}$ & 0.982 & -1.479 & \\
    \cline{2-6}
    & $1\sigma$ range & [1.098, 2.053] & [0.006, 0.007] & [0.979, 0.984] & [-1.482, -1.478]& 0.636\\
    \cline{1-6}
     \multirow{2}{*}{$\tau$ mode}
    & Central value & 0.966 & 0.284 & 0.105 & -0.607 & \\
     \cline{2-6}
    & $1\sigma$ range & [0.649, 1.392] & [0.262, 0.291] & [-0.035, 0.279] & [-0.711, -0.525]& [0.586, 0.688]\\
    \hline
    \hline
    \hline
     $B_s \to K^* l \nu$&  
    &$BR\times 10^{-4}$ &$\langle A_{FB}^{l} \rangle$& $\langle P^{l} \rangle$&$\langle C_{F}^{l} \rangle$&$R_{{B_s}K^{\ast}} $ \\
    \hline
    \hline
    \multirow{2}{*}{$\mu$ mode}
    & Central value & 3.259 & -0.281 & 0.993 & -0.417 & \\
    \cline{2-6}
    & $1\sigma$ range & [2.501, 4.179] & [-0.342, -0.222] & [0.989, 0.995] & [-0.575, -0.247]& 0.578\\
    \cline{1-6}
     \multirow{2}{*}{$\tau$ mode}
    & Central value & 1.884 & -0.132 & 0.539 & -0.105 & \\
     \cline{2-6}
    & $1\sigma$ range & [1.449, 2.419] & [-0.203, -0.061] & [0.458, 0.603] & [-0.208, -0.007]& [0.539, 0.623]\\
    \hline
    \hline
     \hline
    $B \to \pi l \nu$&  
    &$BR \times 10^{-4}$ &$\langle A_{FB}^{l} \rangle$& $\langle P^{l} \rangle$&$\langle C_{F}^{l} \rangle$&$R_{\pi}$ \\
    \hline
    \hline
    \multirow{2}{*}{$\mu$ mode}
    & Central value & 1.369 & $4.678\times 10^{-3}$ & 0.988 & -1.486 & \\
    \cline{2-6}
    & $1\sigma$ range & [1.030, 1.786] & [0.004, 0.006] & [0.981, 0.991] & [-1.489, -1.481]& 0.641\\
    \cline{1-6}
     \multirow{2}{*}{$\tau$ mode}
    & Central value & 0.878 & 0.246 & 0.298 & -0.737 & \\
     \cline{2-6}
    & $1\sigma$ range & [0.690, 1.092] & [0.227, 0.262] & [0.195, 0.385] & [-0.781, -0.682]& [0.576, 0.725]\\
    \hline
    \hline
\end{tabular}}
\caption{The central values and $1\sigma$ ranges of each observable for both $\mu$ and $\tau$ modes in SM are reported
for $B_s \to K l \nu$, $B_s \to K^* l \nu$ and $B \to \pi l \nu$ decays.}
\label{1sigval}
\end{table}

\subsection{Beyond the SM predictions}

\begin{figure}[htbp]
\centering
\includegraphics[width=4.0cm,height=2.3cm]{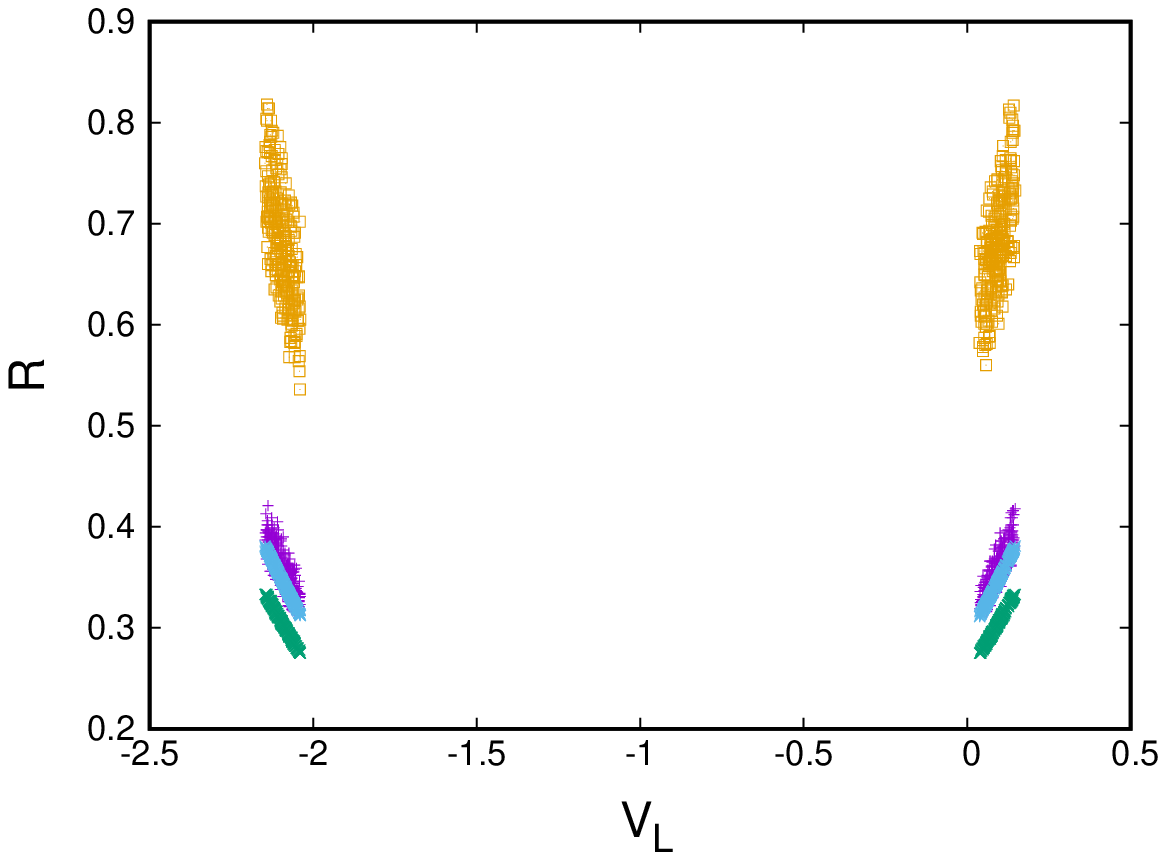}
\includegraphics[width=4.0cm,height=2.3cm]{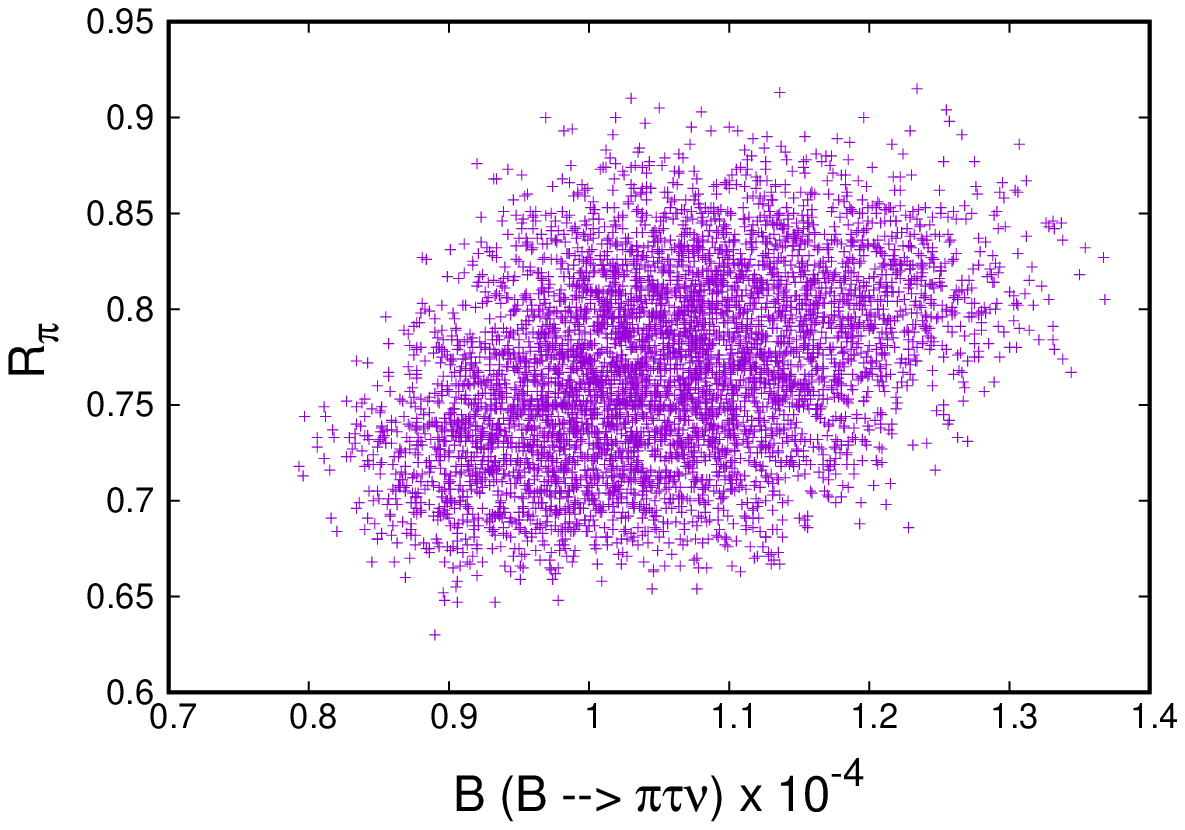}
\includegraphics[width=4.0cm,height=2.3cm]{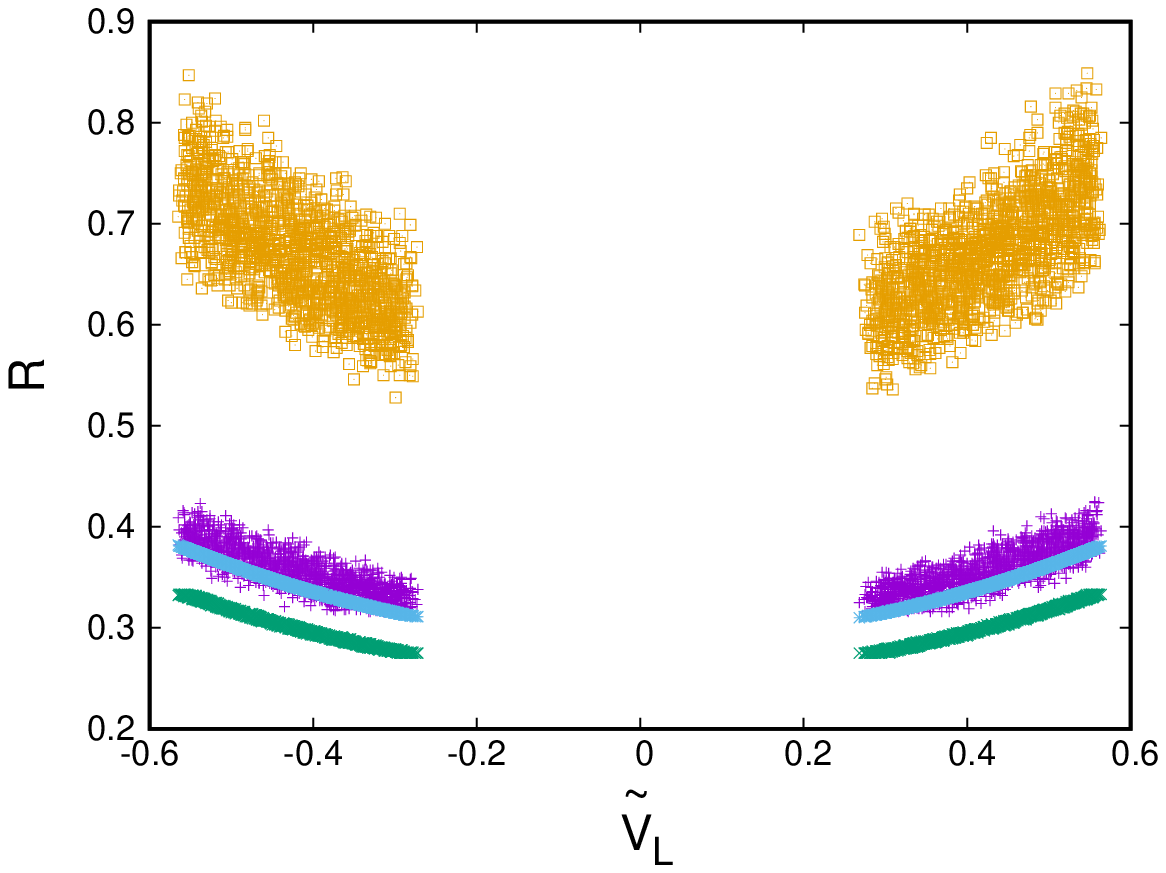}
\includegraphics[width=4.0cm,height=2.3cm]{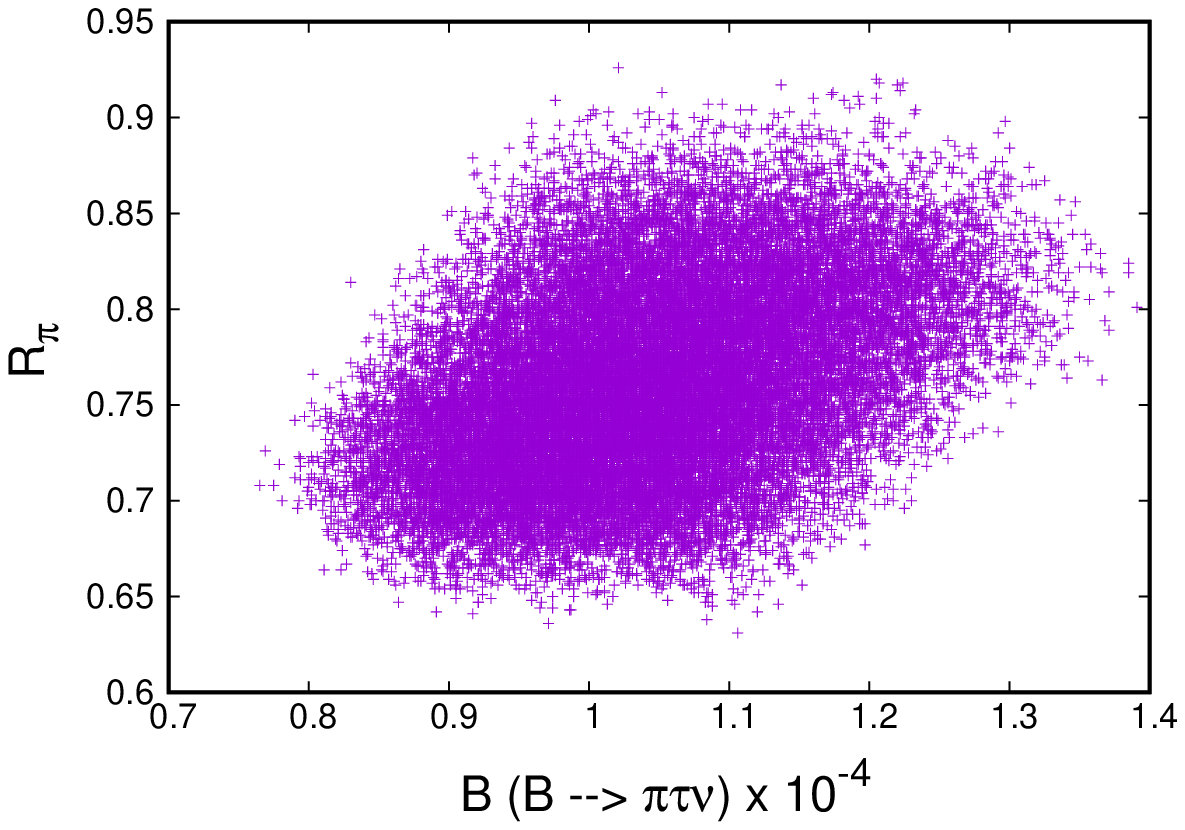}
\caption{In the left panel we show the allowed ranges in $V_L$ (above) and $\widetilde{V}_L$ (below) NP coupling and the corresponding ranges 
in $R_D$~(violet), $R_{D^{\ast}}$~(green), $R_{J/\Psi}$~(blue), and $R_{\pi}^l$~(yellow) once $2\sigma$ experimental constraint is imposed. 
The corresponding ranges in $\mathcal B(B \to \pi\tau\nu)$ and $R_{\pi}$ are shown in the right panel.}
\label{figconvl}
\end{figure}

\begin{figure}[htbp]
\centering
\includegraphics[width=3.6cm,height=2.3cm]{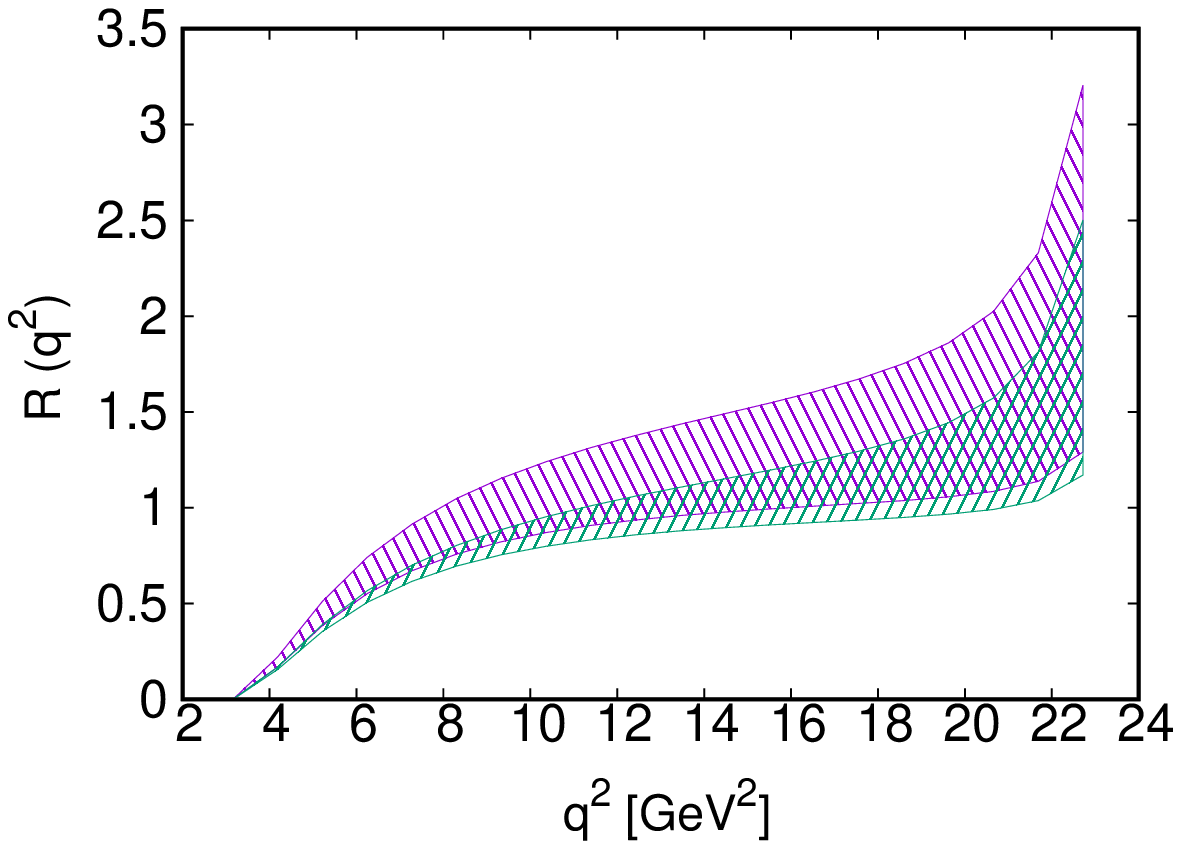}
\includegraphics[width=3.6cm,height=2.3cm]{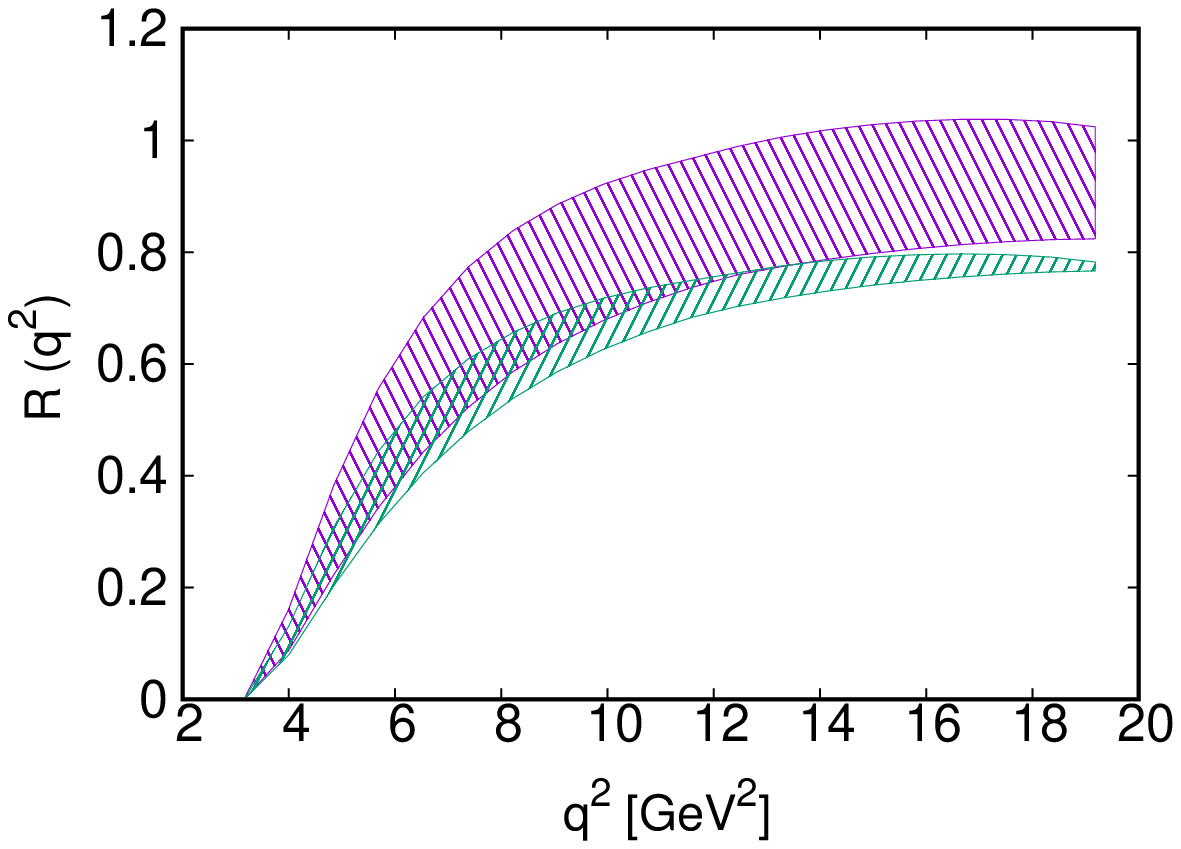}
\includegraphics[width=3.6cm,height=2.3cm]{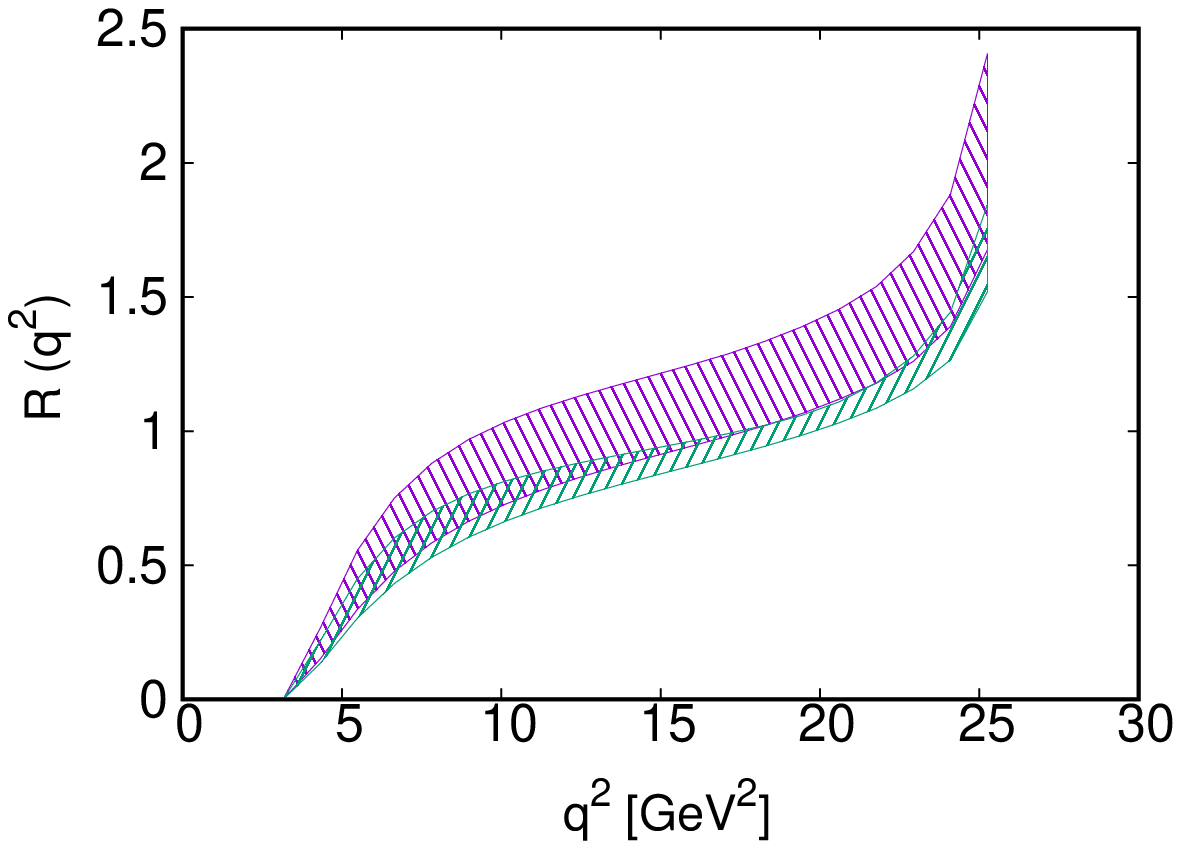}
\includegraphics[width=3.6cm,height=2.3cm]{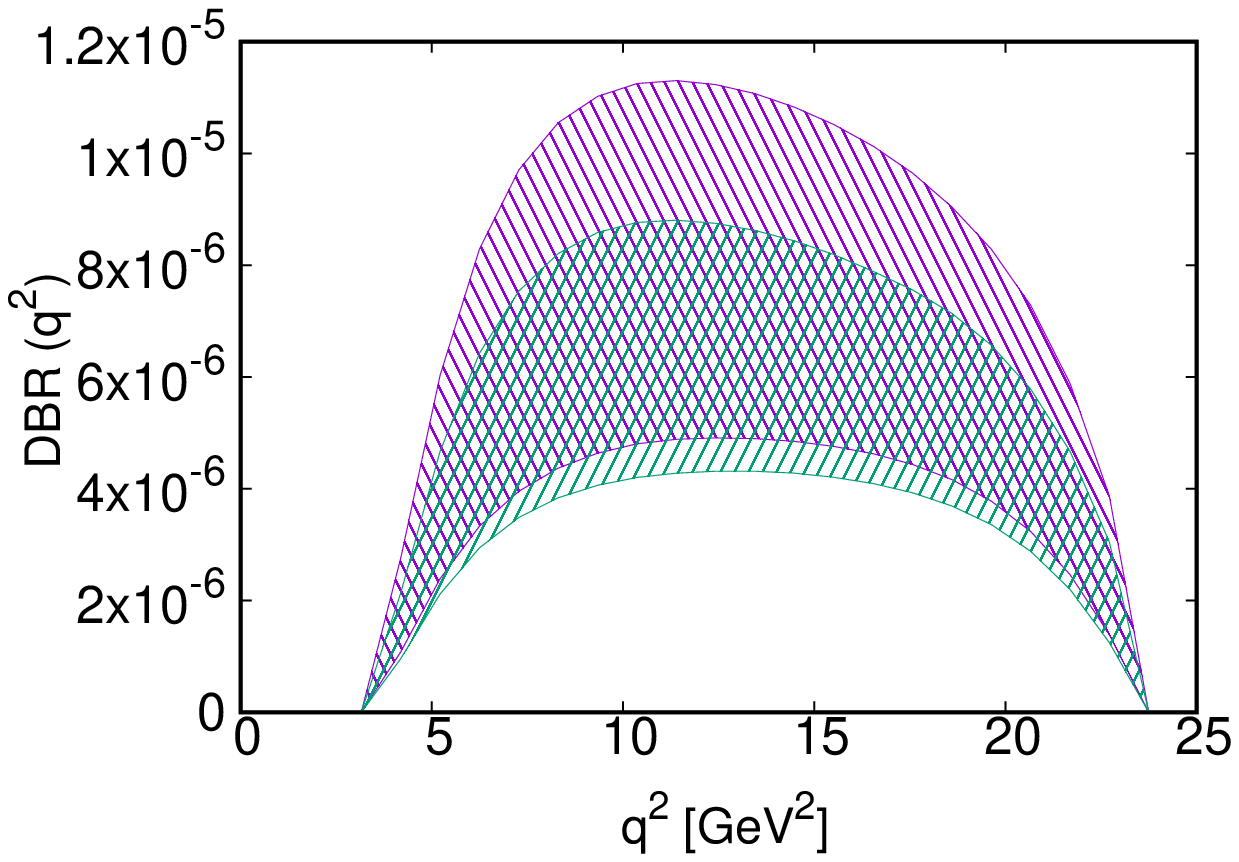}
\includegraphics[width=3.6cm,height=2.3cm]{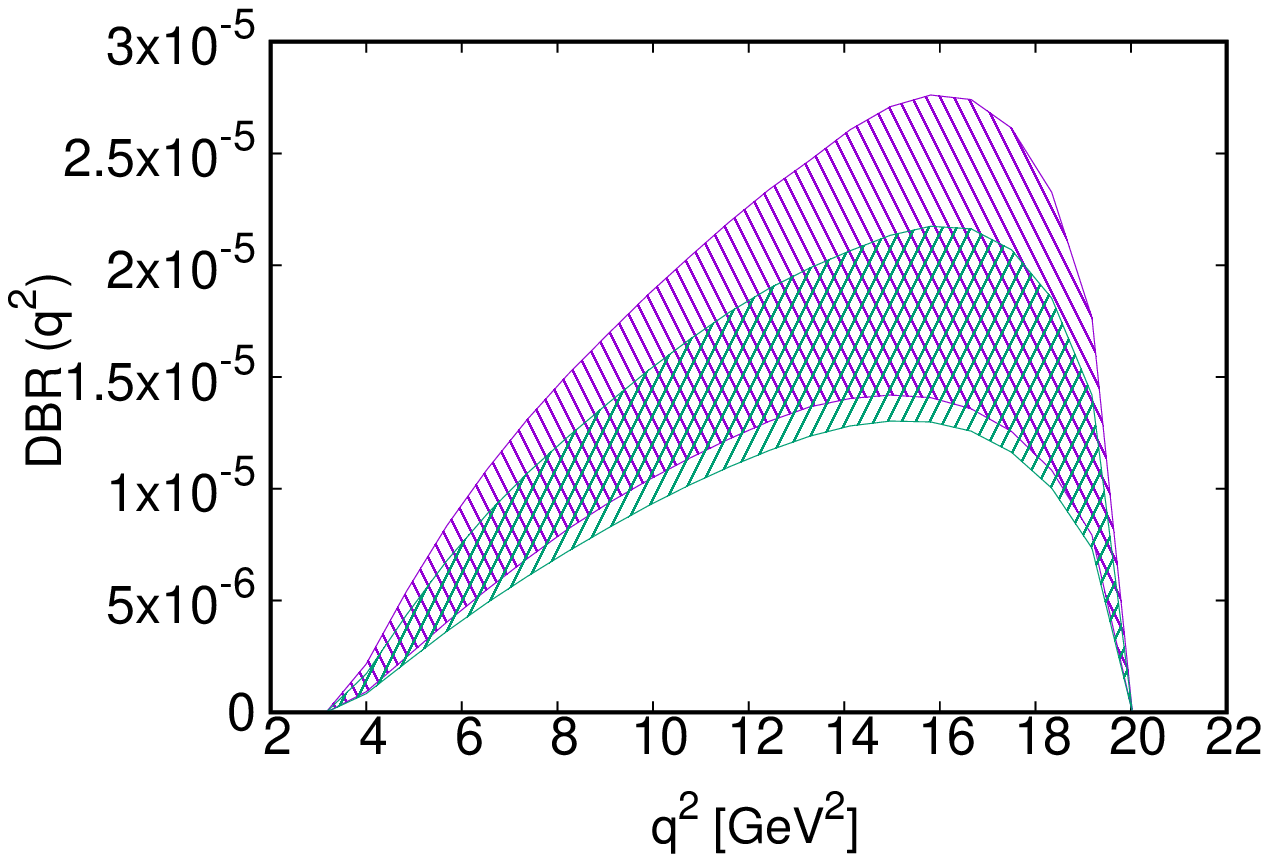}
\includegraphics[width=3.6cm,height=2.3cm]{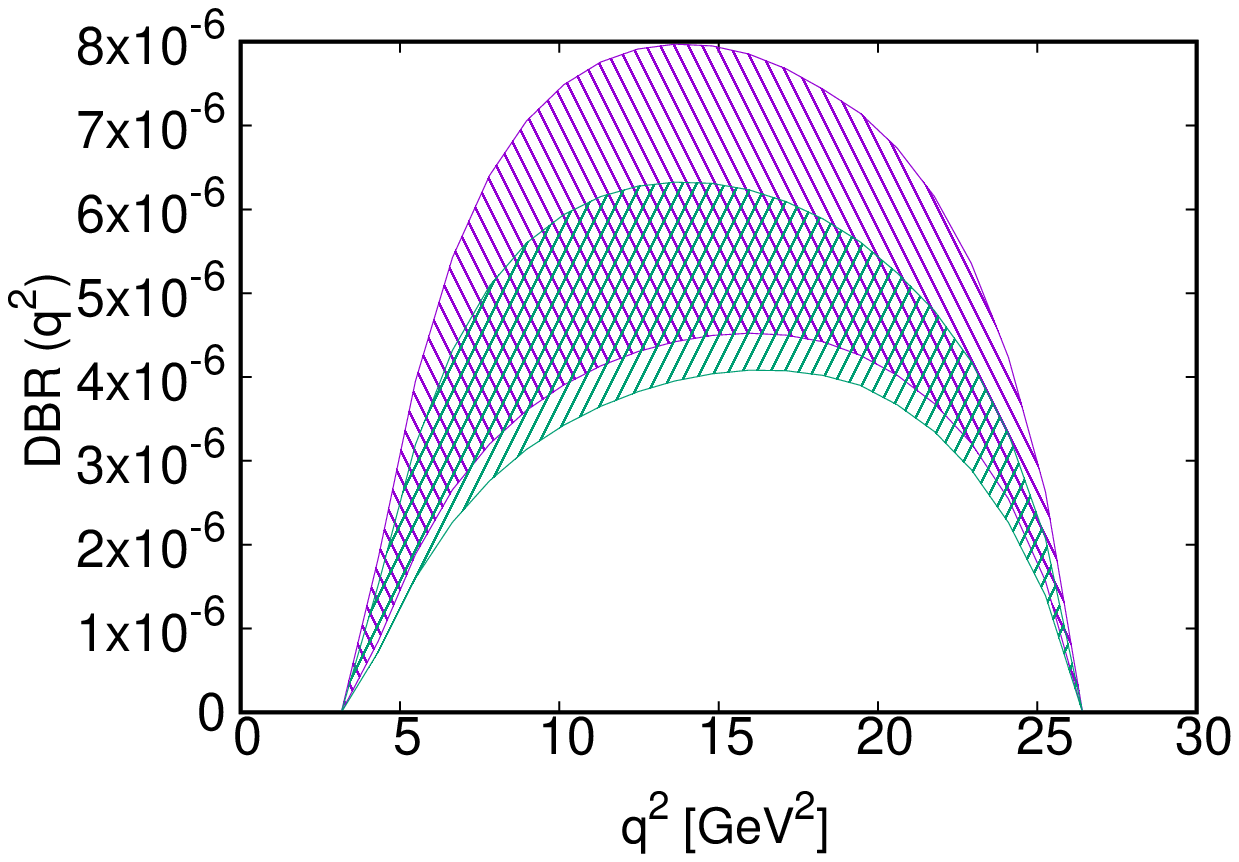}
\caption{$R(q^2)$ and ${\rm DBR}(q^2)$ for  
$B_s \to K \tau \nu$~(first column), $B_s \to K^{\ast} \tau \nu$~(second column) and $B \to \pi \tau \nu$~(third column) decays 
using the $V_L$ NP coupling of Fig.~\ref{figconvl} are shown with violet band. The corresponding SM ranges are shown with green band}
\label{figvl}
\end{figure}

\begin{figure}[htbp]
\centering
\includegraphics[width=3.6cm,height=2.3cm]{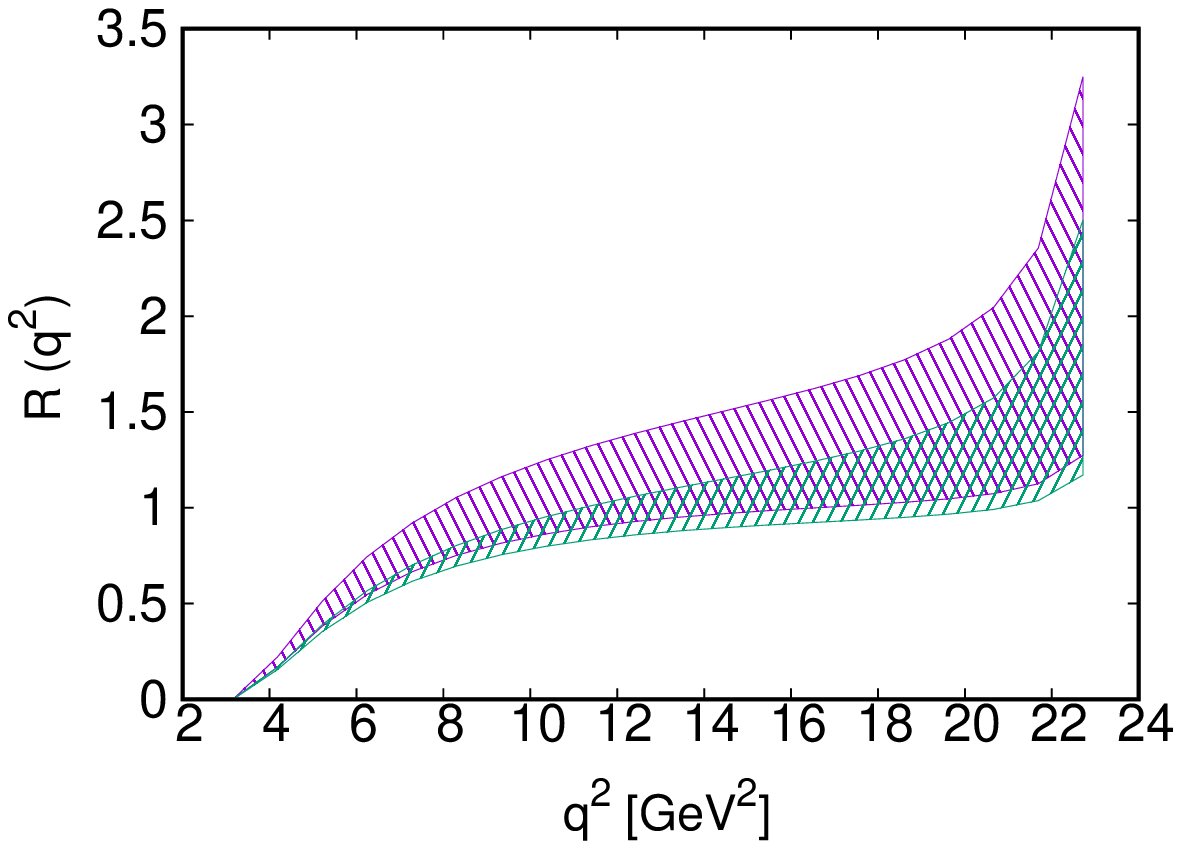}
\includegraphics[width=3.6cm,height=2.3cm]{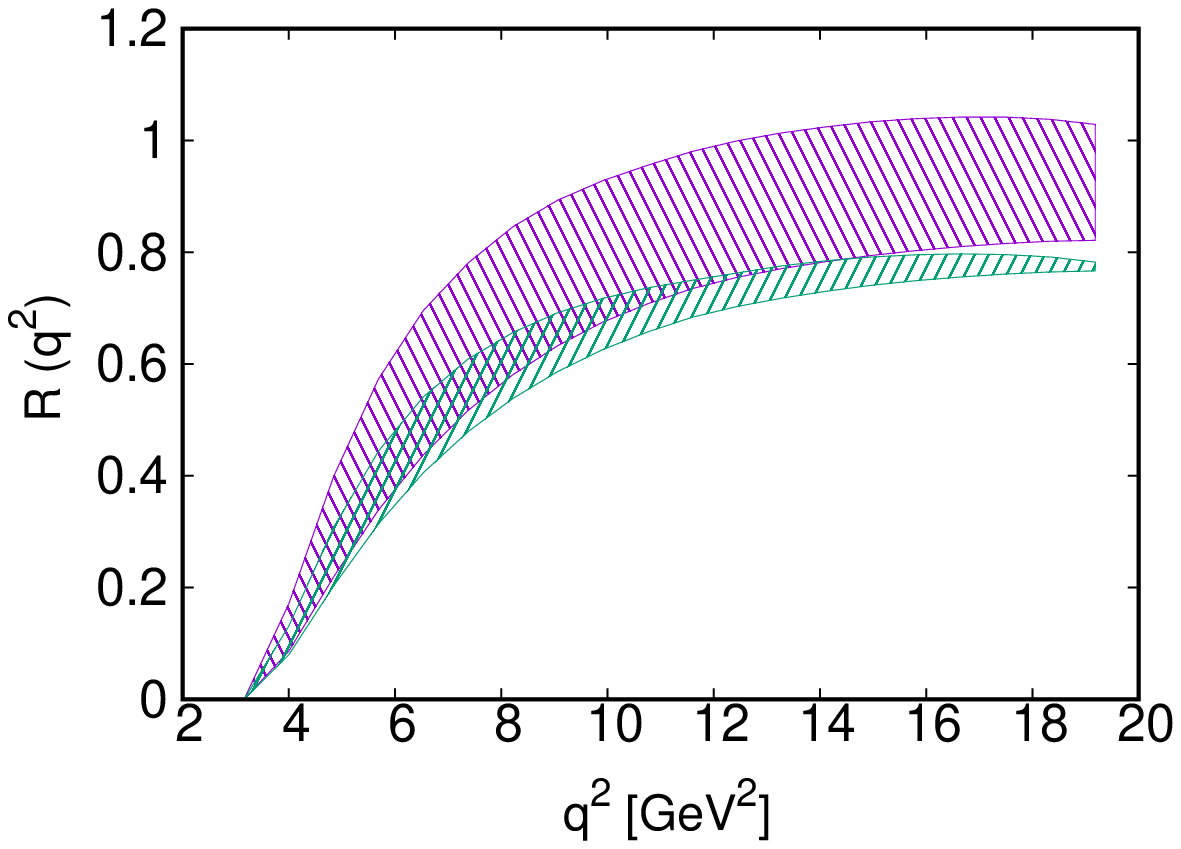}
\includegraphics[width=3.6cm,height=2.3cm]{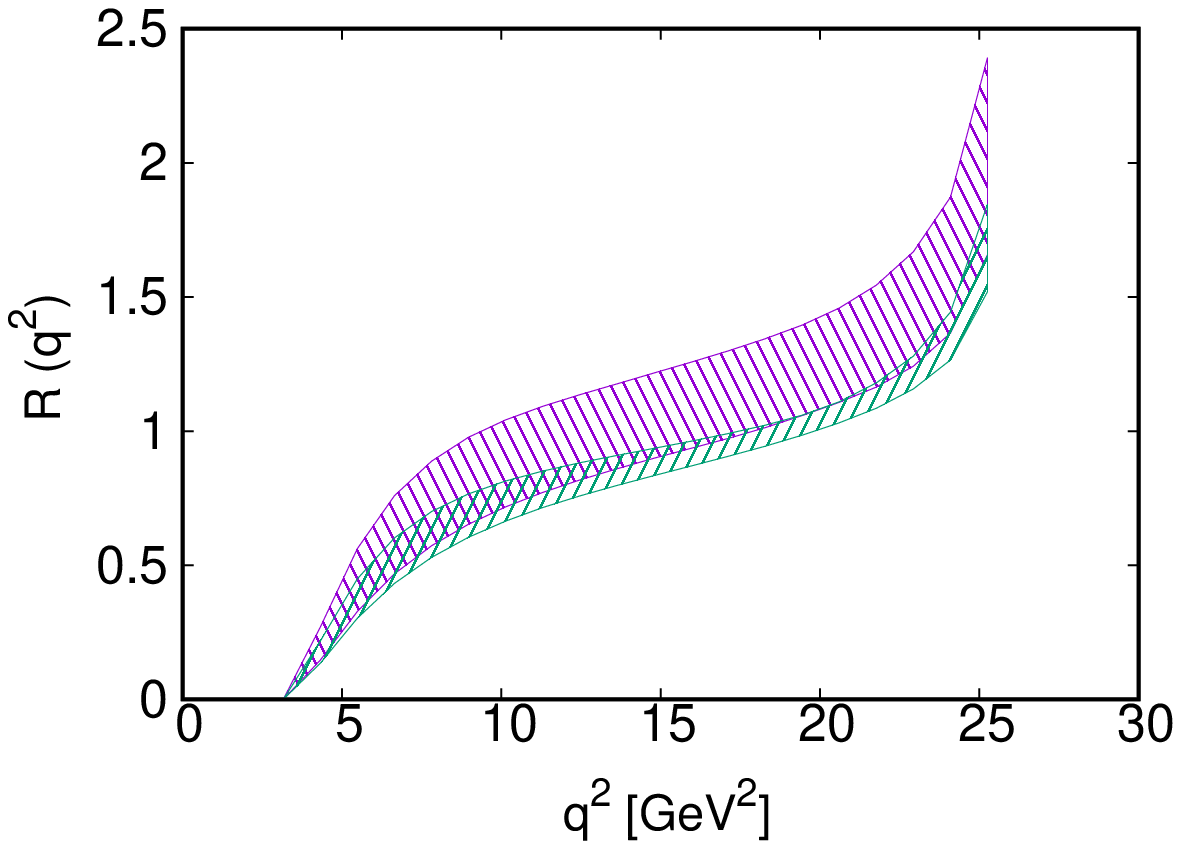}
\includegraphics[width=3.6cm,height=2.3cm]{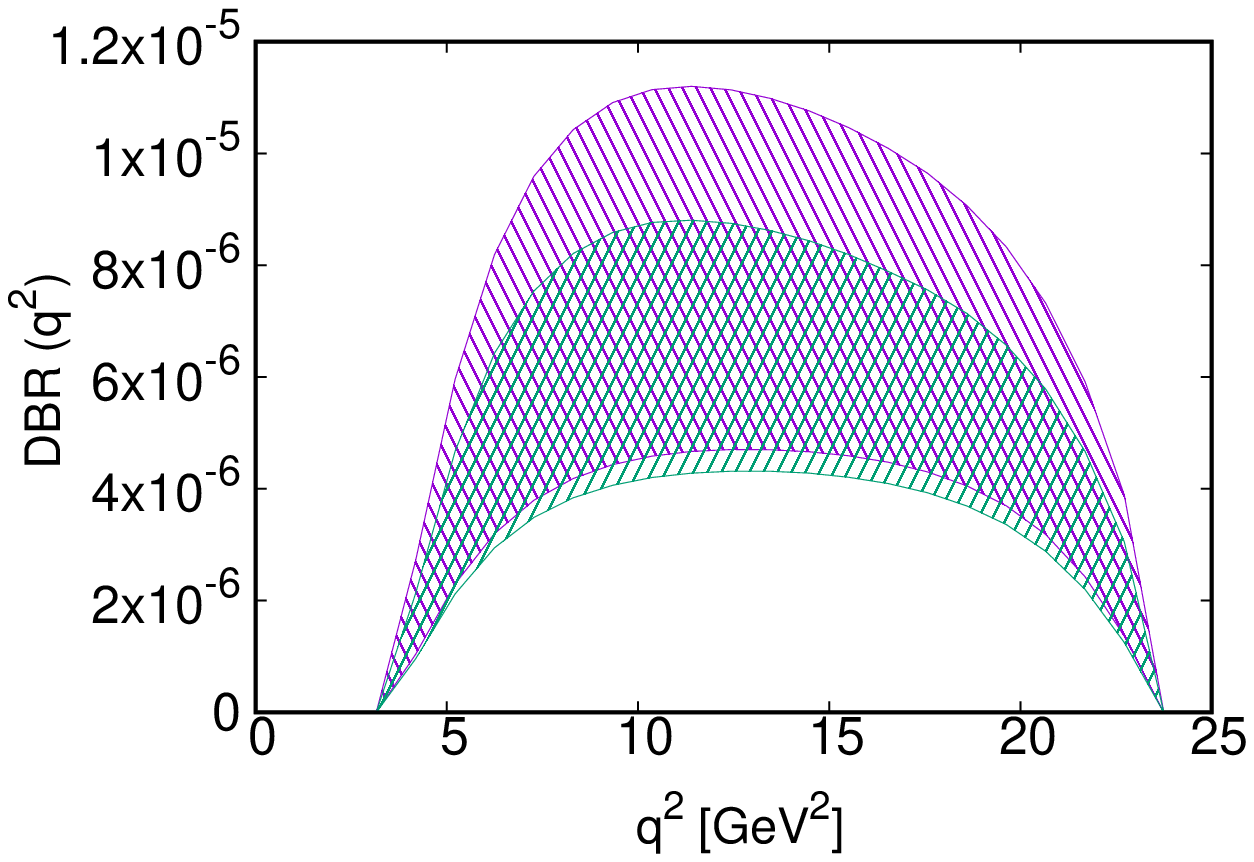}
\includegraphics[width=3.6cm,height=2.3cm]{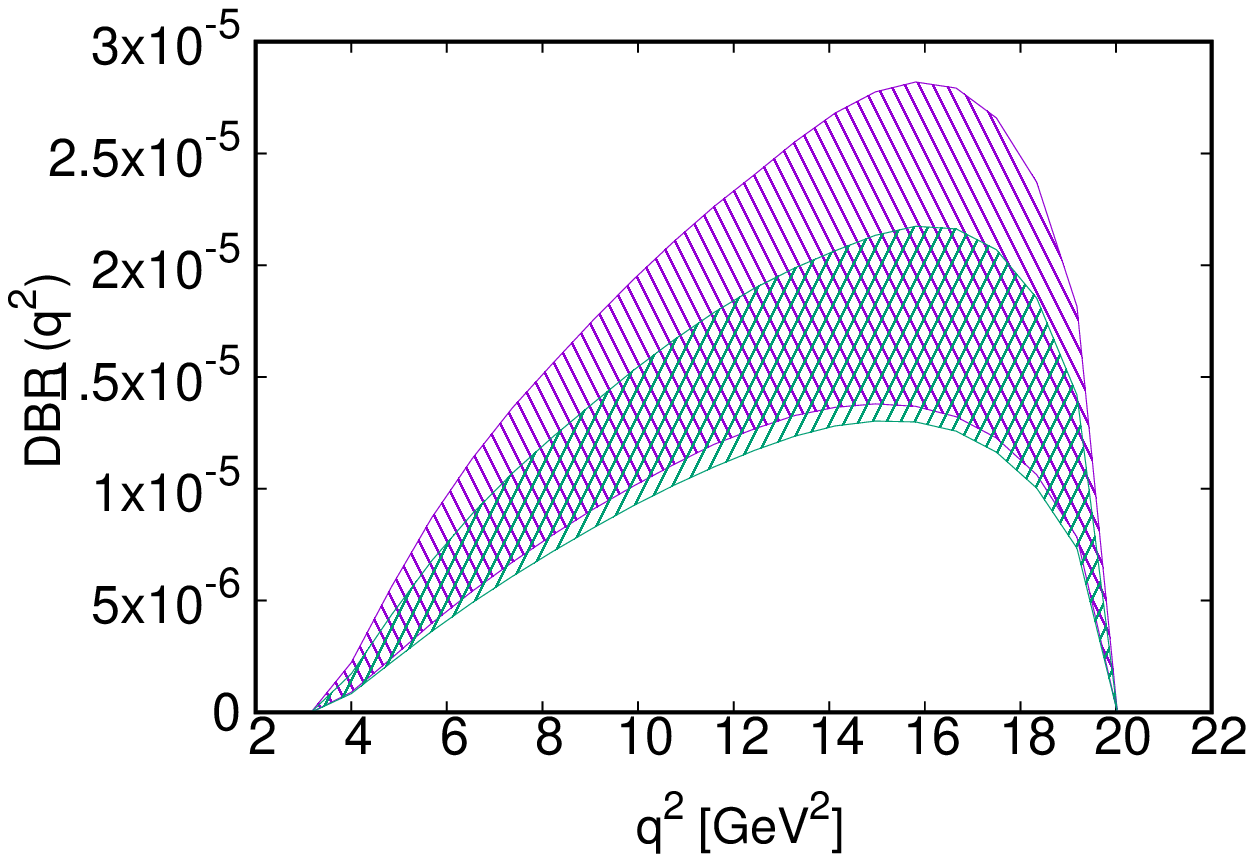}
\includegraphics[width=3.6cm,height=2.3cm]{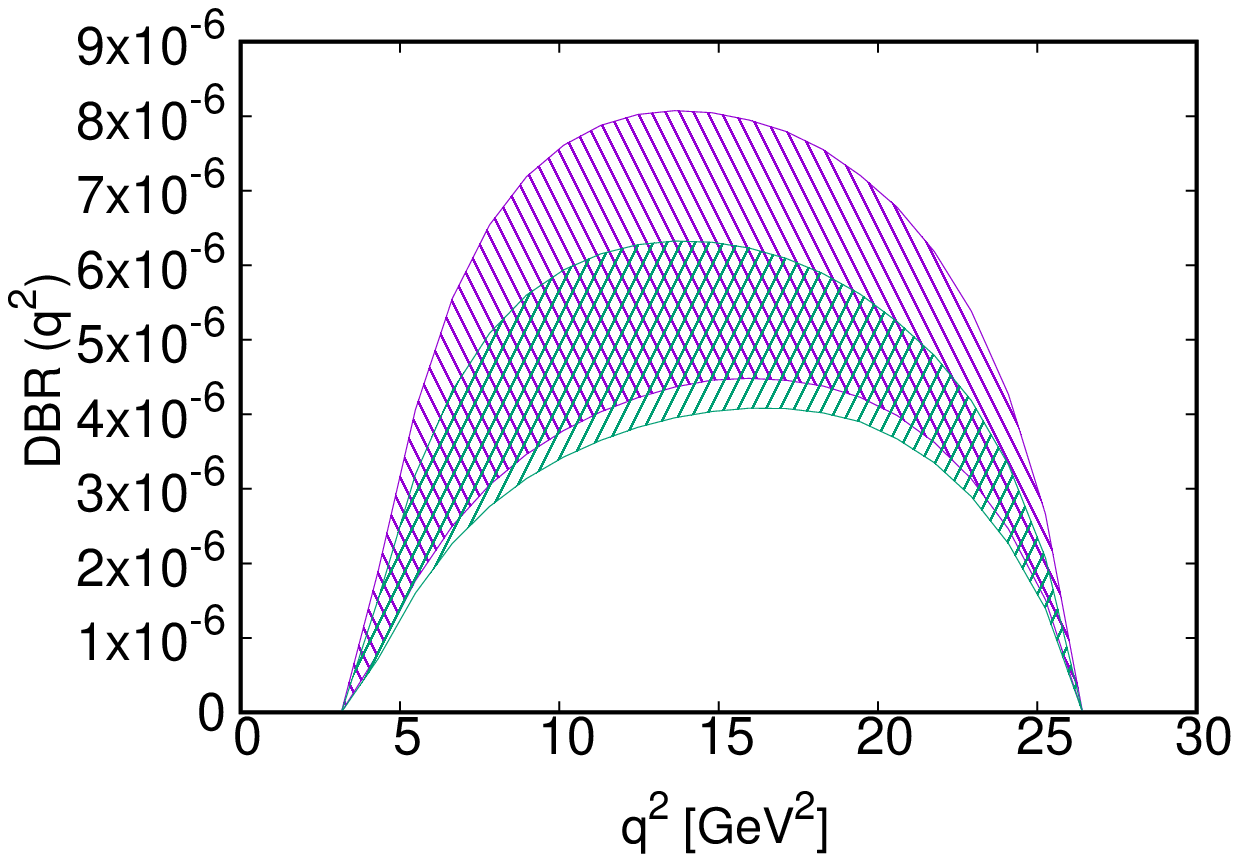}
\includegraphics[width=3.6cm,height=2.3cm]{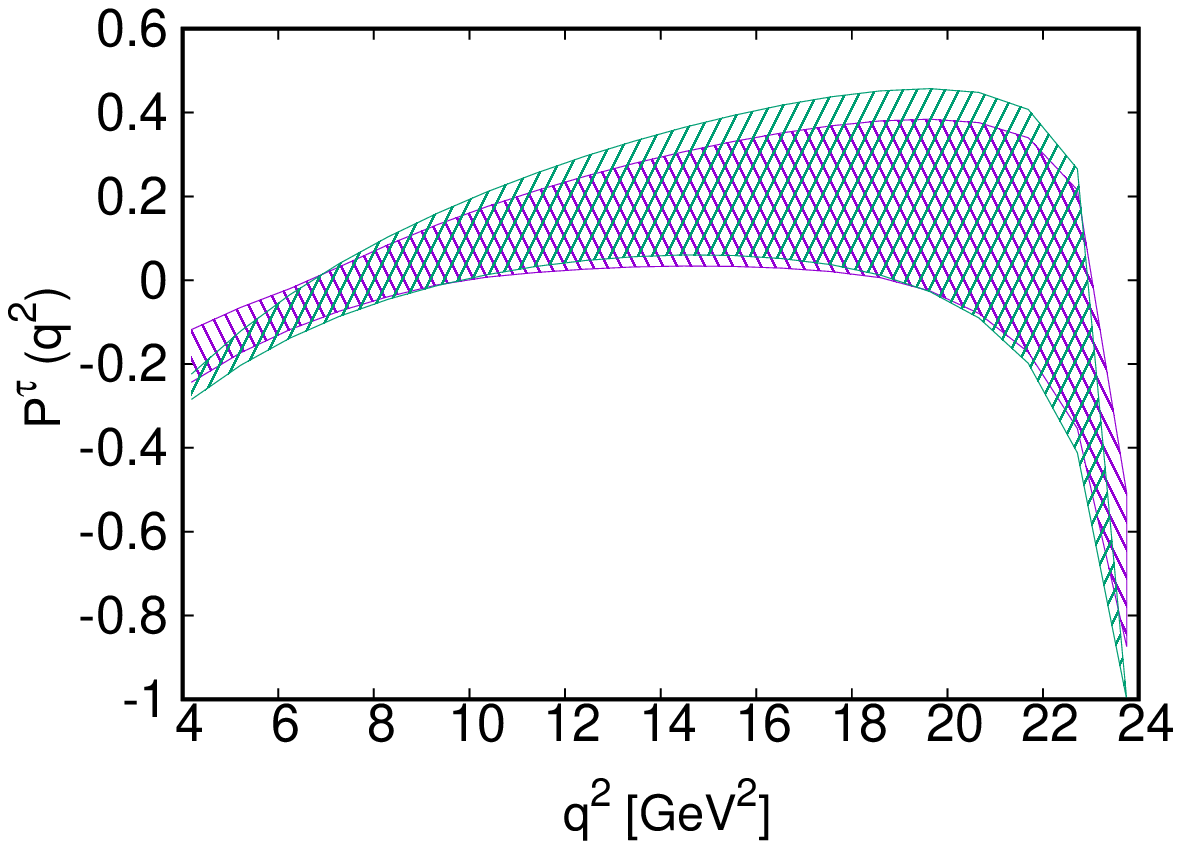}
\includegraphics[width=3.6cm,height=2.3cm]{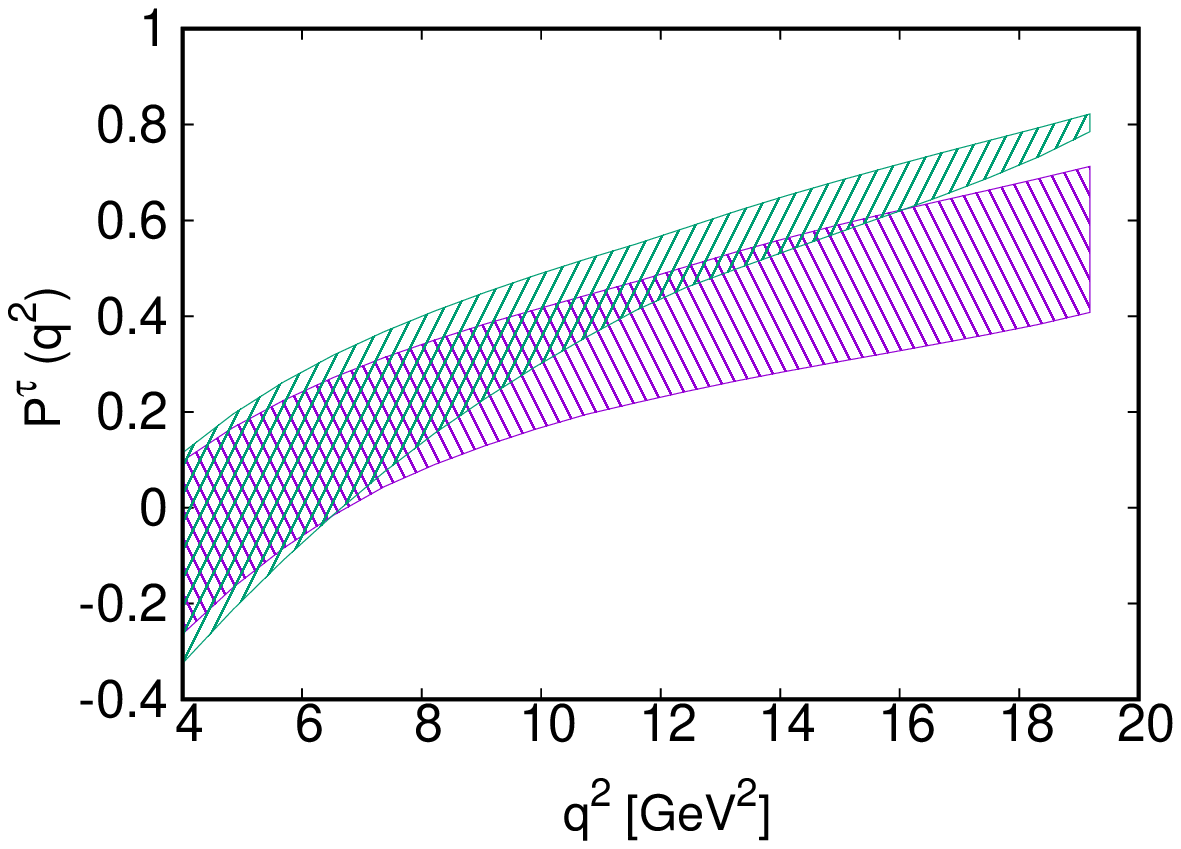}
\includegraphics[width=3.6cm,height=2.3cm]{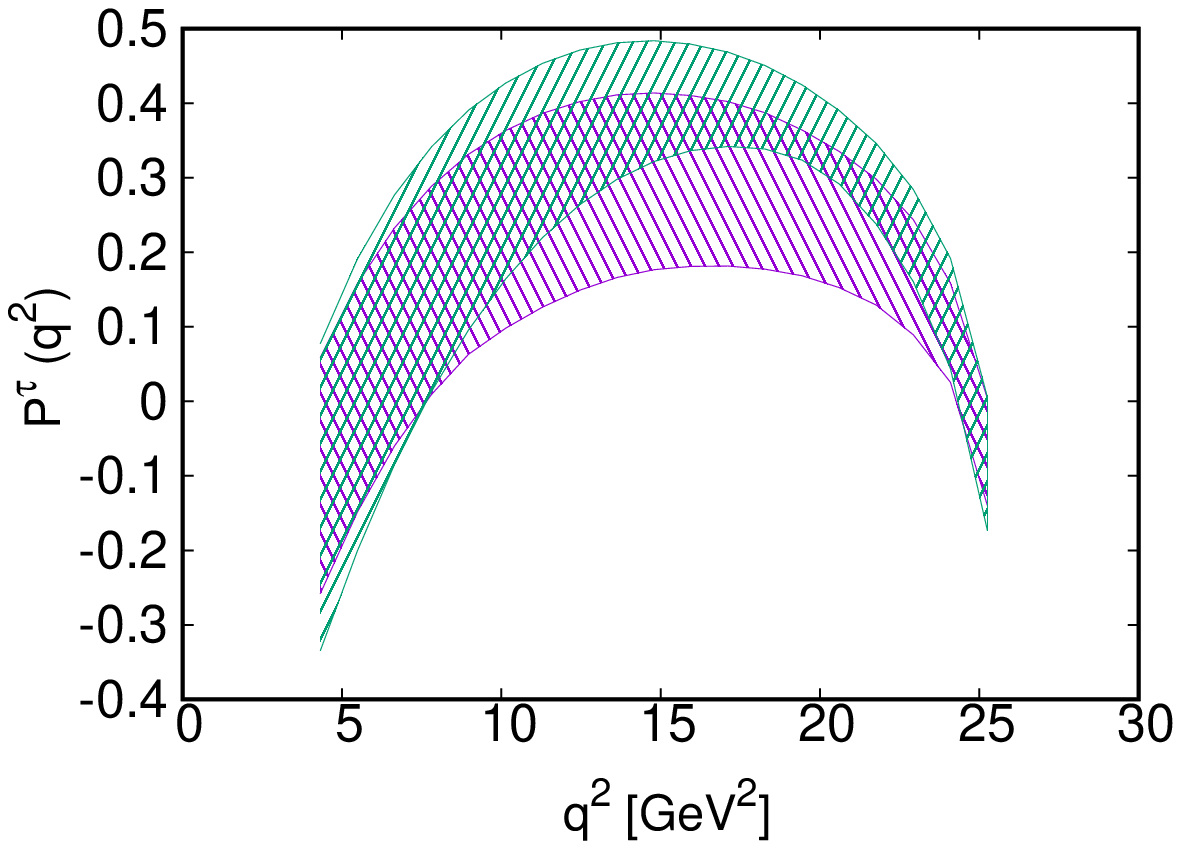}
\caption{{$R(q^2)$, ${\rm DBR}(q^2)$ and $P^{\tau}(q^2)$ for
$B_s \to K \tau \nu$~(first column), $B_s \to K^{\ast} \tau \nu$~(second column) and $B \to \pi \tau \nu$~(third column) decays
using the $\widetilde{V}_L$ NP coupling is shown in violet band. The corresponding $1\sigma$ SM band is shown 
in green color.}}
\label{figvlt}
\end{figure}

We discuss the NP contributions coming from $V_L$ and $\widetilde{V}_L$ NP couplings. To get the allowed NP parameter space, 
we impose $2\sigma$ constraint coming from the measured values of $R_D$, $R_{D^{\ast}}$, $R_{J/\Psi}$, and $R_{\pi}^l$. In the left panel of
Fig.~\ref{figconvl}, we show the allowed range of $V_L$ and $\widetilde{V}_L$ NP couplings once the $2\sigma$ constraints are imposed. 
Similarly, in the right panel the corresponding ranges in
$\mathcal B(B \to \pi\tau\nu)$ and $R_{\pi}$ using the allowed ranges of $V_L$ and $\widetilde{V}_L$ NP couplings are shown.
In Table~\ref{vlrang} we display the allowed ranges of each observable in the presence of $V_L$ and $\widetilde{V}_L$ NP couplings.
Also, in Fig.~\ref{figvl} and ~\ref{figvlt}, we display the $q^2$ dependency of the various observables in the presence of $V_L$ and
$\widetilde{V}_L$ NP couplings for the $B_s \to K \tau \nu$, $B_s \to K^{\ast}\tau \nu$ and $B \to \pi \tau \nu$ decays. 
The detailed observations are as follows:

\begin{itemize}
 \item For the $V_L$ NP coupling, we notice a significant deviation from the SM prediction in ${\rm DBR}(q^2)$ and $R(q^2)$ 
 for all the decay modes. In addition, in the presence of $\widetilde{V}_L$ NP coupling the $\tau$ polarization fraction show deviation along 
 with $R(q^2)$ and ${\rm DBR}(q^2)$. So the measurement of $P^{\tau}(q^2)$ can easily differentiate $V_L$ and $\widetilde{V}_L$ NP contributions.
 \item The other observable such as $A_{FB}^{\tau}(q^2)$, $P^{\tau}(q^2)$ and $C_{F}^{\tau}(q^2)$ are not affected by $V_L$ NP coupling.
 Similarly, $A_{FB}^{\tau}(q^2)$ and $C_{F}^{\tau}(q^2)$ are not affected by $\widetilde{V}_L$ NP coupling.
 \end{itemize}

\begin{table}[htbp]
\centering
\begin{tabular}{|c|c|c||c|c|c|}
    \hline
    &\multicolumn{2}{c|}{$V_L$ } &\multicolumn{3}{c|}{$\widetilde{V}_L$} \\
    \cline{2-6}
    &$\langle R \rangle$ & $\langle BR \rangle \times 10^{-4}$ & $\langle R \rangle$& $\langle BR \rangle \times 10^{-4}$ &$\langle P^{\tau} \rangle$ \\
    \hline
    \hline
    $B_s \to K \tau \nu$ & $[0.644, 0.891]$ & $[0.735, 1.746]$ & $[0.638, 0.898]$ & $[0.731, 1.774]$ & $[-0.026, 0.217]$ \\
    \hline
    $B_s \to K^{\ast} \tau \nu$ & $[0.593, 0.804]$ & $[1.684, 2.993]$ & $[0.582, 0.802]$ & $[1.579, 3.098]$ & $[0.249, 0.513]$ \\
    \hline
    $B \to \pi \tau \nu$ & $[0.630, 0.915]$ & $[0.793, 1.368]$ & $[0.631, 0.926]$ & $[0.765, 1.391]$ & $[0.117, 0.315]$ \\
    \hline
\end{tabular}
\caption{Allowed ranges of each observable in the presence of $V_L$ and $\widetilde{V}_L$ NP coupling.}
\label{vlrang}
\end{table}
  
\section{Conclusion}
We study $B_s \to (K,\,K^{\ast})\tau\nu$ and $B \to \pi\tau\nu$ decay modes within the SM and within the various NP scenarios. 
Although, there are hints of NP in various $B$ meson decays, the NP is not yet established. 
Studying $B_s \to (K,\,K^{\ast})\tau\nu$ and $B \to \pi\tau\nu$ 
decay modes theoretically as well as experimentally are well motivated as these can provide complementary information regarding NP.


\begin{thebibliography}{99}
  
  \bibitem{Lattice:2015rga} 
  J.~A.~Bailey {\it et al.} [MILC Collaboration],
  Phys.\ Rev.\ D {\bf 92}, no. 3, 034506 (2015)
  
  \bibitem{Na:2015kha} 
  H.~Na {\it et al.} [HPQCD Collaboration],
  Phys.\ Rev.\ D {\bf 92}, no. 5, 054510 (2015)
  Erratum: [Phys.\ Rev.\ D {\bf 93}, no. 11, 119906 (2016)]
  
  \bibitem{Aoki:2016frl} 
  S.~Aoki {\it et al.},
  Eur.\ Phys.\ J.\ C {\bf 77}, no. 2, 112 (2017)

  \bibitem{Bigi:2016mdz} 
  D.~Bigi and P.~Gambino,
  Phys.\ Rev.\ D {\bf 94}, no. 9, 094008 (2016)
  
  
  
   \bibitem{Fajfer:2012vx} 
  S.~Fajfer, J.~F.~Kamenik and I.~Nisandzic,
  Phys.\ Rev.\ D {\bf 85}, 094025 (2012)
  
  \bibitem{Bernlochner:2017jka} 
  F.~U.~Bernlochner, Z.~Ligeti, M.~Papucci and D.~J.~Robinson,
  Phys.\ Rev.\ D {\bf 95}, no. 11, 115008 (2017)
  Erratum: [Phys.\ Rev.\ D {\bf 97}, no. 5, 059902 (2018)]
  
  \bibitem{Bigi:2017jbd} 
  D.~Bigi, P.~Gambino and S.~Schacht,
  JHEP {\bf 1711}, 061 (2017)
  
  \bibitem{Jaiswal:2017rve} 
  S.~Jaiswal, S.~Nandi and S.~K.~Patra,
  JHEP {\bf 1712}, 060 (2017)
  
   \bibitem{Cohen:2018dgz} 
  T.~D.~Cohen, H.~Lamm and R.~F.~Lebed,
  arXiv:1807.02730 [hep-ph].  
  
  \bibitem{Bona:2009cj} 
  M.~Bona {\it et al.} [UTfit Collaboration],
  Phys.\ Lett.\ B {\bf 687}, 61 (2010)
  
  
   \bibitem{Patrignani:2016xqp} 
  C.~Patrignani {\it et al.} [Particle Data Group],
  Chin.\ Phys.\ C {\bf 40}, no. 10, 100001 (2016).
    
  \bibitem{Lees:2013uzd} 
  J.~P.~Lees {\it et al.} [BaBar Collaboration],
  Phys.\ Rev.\ D {\bf 88}, no. 7, 072012 (2013)
  
  \bibitem{Huschle:2015rga} 
  M.~Huschle {\it et al.} [Belle Collaboration],
  Phys.\ Rev.\ D {\bf 92}, no. 7, 072014 (2015)
  
    \bibitem{Sato:2016svk} 
  Y.~Sato {\it et al.} [Belle Collaboration],
  Phys.\ Rev.\ D {\bf 94}, no. 7, 072007 (2016)
  
  \bibitem{Hirose:2016wfn} 
  S.~Hirose {\it et al.} [Belle Collaboration],
  Phys.\ Rev.\ Lett.\  {\bf 118}, no. 21, 211801 (2017)
  
  \bibitem{Aaij:2015yra} 
  R.~Aaij {\it et al.} [LHCb Collaboration],
  Phys.\ Rev.\ Lett.\  {\bf 115}, no. 11, 111803 (2015)
  Erratum: [Phys.\ Rev.\ Lett.\  {\bf 115}, no. 15, 159901 (2015)]
   
   \bibitem{Aaij:2017tyk} 
  R.~Aaij {\it et al.} [LHCb Collaboration],
  arXiv:1711.05623 [hep-ex].
  
    \bibitem{Bernlochner:2015mya} 
  F.~U.~Bernlochner,
  Phys.\ Rev.\ D {\bf 92}, no. 11, 115019 (2015)
  
  
   \bibitem{Dutta:2013qaa} 
  R.~Dutta, A.~Bhol and A.~K.~Giri,
  Phys.\ Rev.\ D {\bf 88}, no. 11, 114023 (2013)
  

  
  \bibitem{Rajeev:2018txm} 
  N.~Rajeev and R.~Dutta,
  Phys.\ Rev.\ D {\bf 98}, no. 5, 055024 (2018)
  
   
  
\end{thebibliography}
\end{document}